\documentclass[prx,twocolumn,english,superscriptaddress,floatfix,longbibliography]{revtex4-2}

\usepackage{graphicx}
\usepackage{txfonts}
\usepackage{dcolumn}   
\usepackage{bm}
\usepackage{amssymb}  
\usepackage{amssymb} 
\usepackage{verbatim}
\usepackage{multirow}
\usepackage{xcolor} 
\usepackage{braket}
\usepackage{amsfonts}
\usepackage{soul}
\usepackage{hyperref}
\hypersetup{colorlinks=true, pdfstartview=FitV, linkcolor=blue, citecolor=black, plainpages=false, urlcolor=blue}
\usepackage[all]{hypcap}

\begin{document}
\title{Quasiparticle poisoning of superconducting qubits with active gamma irradiation}
\author{C. P. Larson}
\affiliation{Department of Physics, Syracuse University, Syracuse, New York 13244-1130}
\author{E. Yelton}
\affiliation{Department of Physics, Syracuse University, Syracuse, New York 13244-1130}
\author{K. Dodge}
\affiliation{Intelligence Community Postdoctoral Research Fellowship Program, Department of Physics, Syracuse University, Syracuse, NY 13244-1130, USA}
\author{K. Okubo}
\affiliation{Department of Physics, Syracuse University, Syracuse, New York 13244-1130}
\author{J. Batarekh}
\affiliation{Department of Physics, Syracuse University, Syracuse, New York 13244-1130}
\affiliation{Department of Physics, University of Wisconsin-Madison, Madison, Wisconsin 53706, USA}
\author{V. Iaia}
\affiliation{Department of Physics, Syracuse University, Syracuse, New York 13244-1130}
\author{N. A. Kurinsky}
\affiliation{Kavli Institute for Particle Astrophysics and Cosmology, \\SLAC National Accelerator Laboratory Menlo Park, CA 94025, USA}
\author{B. L. T. Plourde}
\affiliation{Department of Physics, Syracuse University, Syracuse, New York 13244-1130}
\affiliation{Department of Physics, University of Wisconsin-Madison, Madison, Wisconsin 53706, USA}

\date{\today}

\begin{abstract}
When a high-energy particle, such as a $\gamma$-ray or muon, impacts the substrate of a superconducting qubit chip, large numbers of electron-hole pairs and phonons are created. The ensuing dynamics of the electrons and holes changes the local offset-charge environment for qubits near the impact site. The phonons that are produced have energy above the superconducting gap in the films that compose the qubits, leading to quasiparticle excitations above the superconducting ground state when the phonons impinge on the qubit electrodes. An elevated density of quasiparticles degrades qubit coherence, leading to errors in qubit arrays. Because these pair-breaking phonons spread throughout much of the chip, the errors can be correlated across a large portion of the array, posing a significant challenge for quantum error correction. In order to study the dynamics of $\gamma$-ray impacts on superconducting qubit arrays, we use a $\gamma$-ray source outside the dilution refrigerator to controllably irradiate our devices. By using charge-sensitive transmon qubits, we can measure both the offset-charge shifts and quasiparticle poisoning due to the $\gamma$ irradiation at different doses. We study correlations between offset-charge shifts and quasiparticle poisoning for different qubits in the array and compare this with numerical modeling of charge and phonon dynamics following a $\gamma$-ray impact. We thus characterize the poisoning footprint of these impacts and quantify the performance of structures for mitigating phonon-mediated quasiparticle poisoning.
\end{abstract}

\maketitle

%%%%%%%%%  Introduction  %%%%%%%%% 
\section{Introduction}
A current major challenge to the implementation of fault-tolerant quantum processors with superconducting qubits comes from the impact of high-energy particles from background radioactivity or cosmic-ray muons~\cite{Vepsalainen2020, Cardani2021, Wilen2021, McEwen2022, Harrington2024, Li2024}. Such an impact deposits a large amount of energy into the device substrate, on average $\sim$100 keV for $\gamma$-rays from environmental background radiation and $\sim$500 keV for muons~\cite{Wilen2021}. Recent work has also highlighted the importance of proton and neutron contributions that occur less often but result in energy depositions of several MeV~\cite{Fowler2024}. Each impact generates many electron-hole ($e^{-}/h^{+}$) pairs close to the impact site that can either recombine immediately or travel hundreds of $\muup$m before becoming trapped by charge defects in the substrate. The subsequent rearrangement of the charge environment modifies the induced offset charge for qubits near the impact site. In addition to the electron-hole pairs, a large number of highly energetic phonons are produced with energy well beyond twice the superconducting gap $\Delta$ of the qubit electrodes \cite{Martinis2021}. When these phonons impinge on the superconducting films of the qubit, they break Cooper pairs with high probability, leading to an enhanced density of dissipative quasiparticles (QPs)  near the Josephson junctions of the qubits; an elevated QP density raises the rate of QP tunneling across the junction and increases qubit decoherence, leading to enhanced error rates. Because these phonons travel efficiently throughout the chip, these errors in the aftermath of a high-energy impact tend to be correlated across the qubit array, posing a significant problem for standard quantum error correction schemes (QEC), such as the surface code~\cite{Fowler2012, GoogleQAI2023, GoogleQAI2024}.

Prior experimental studies of correlated errors from high-energy impacts have primarily relied on random energy depositions from a variety of background sources of radiation in the qubit environment, for example, $\gamma$-rays from radioactive contamination in the lab or muons coming from the upper atmosphere. In this manuscript, we describe experimental measurements of offset-charge and QP poisoning dynamics in qubit arrays using a calibrated source of $\gamma$-rays of known energy. By placing the radioactive source outside the dilution refrigerator that contains the qubits and choosing a source with sufficient energy and activity to penetrate through the cryostat shielding to the sample, we ensure that the response of our qubits is dominated by the $\gamma$-rays from the source over impacts from background sources. In addition, we can vary the distance from source to sample $r$ and thus control the dose rate of $\gamma$ irradiation at the chip [see Fig.~\ref{fig:intro}(a)]. With this approach, we are able to study the dynamics of the poisoning process from $\gamma$-rays of a particular energy and compare directly with Monte Carlo-based numerical modeling of the impact process. Furthermore, we can assess the effectiveness of mitigation techniques for suppressing correlated errors by characterizing a QP poisoning footprint. 

Various approaches have been explored for reducing correlated errors from ionizing impacts on qubit arrays. There have been several efforts to reduce the frequency of ionizing impacts in the qubit chip, including operating the devices in deep underground facilities to reduce the cosmic-ray muon flux~\cite{Cardani2021, Cardani2023, Dedominicis2024, Bratrud2024}, shielding the cryostat with lead to reduce the external $\gamma$-ray flux~\cite{Vepsalainen2020, Cardani2021, Cardani2023, Li2024, Loer2024, Dedominicis2024, Bratrud2024}, and choosing materials for the packaging, cabling, and internal shielding that limits further radiation to the sample~\cite{Cardani2021, Cardani2023, Loer2024}. While these efforts are important, future computations on a quantum processor could run for days or longer; keeping the processor free of ionizing radiation over such a timescale may prove to be exceptionally challenging. At the device level, gap engineering, with sufficiently different superconducting gaps on either side of the Josephson junction \cite{Harrington2024, Mcewen2024}, has been shown to be effective for suppressing tunneling of excess QPs across the junction. Another approach involves channeling the energy in the substrate following an ionizing impact away from the qubits through phonon downconversion~\cite{Iaia2022}. Attenuating the number of pair-breaking phonons in the substrate following an impact leads to fewer phonon scattering events in the qubit junction electrodes, and thus reduced QP poisoning. Increased attenuation for phonon trajectories farther from the impact site is expected to produce a footprint, with the largest enhancement in QP-induced errors in qubits near the impact and progressively less poisoning away from this region \cite{Yelton2024}. QEC schemes capable of responding to transient elevations in error rates in certain regions of a qubit lattice have been developed recently \cite{Sane2023, Xu2022}. For such an approach, knowledge of the poisoning location and footprint is crucial.

Our experiments with controlled $\gamma$ irradiation allow us to identify the location of a $\gamma$-ray impact and study the correlated QP poisoning for all of the qubits in the array. We are thus able to characterize the poisoning footprint for different levels of phonon mitigation and compare with numerical modeling.

%%%%%%%%%  Experimental Approach  %%%%%%%%% 
\section{Experimental Approach}

\begin{figure}[t!]
\centering
\includegraphics[width=8.6cm]{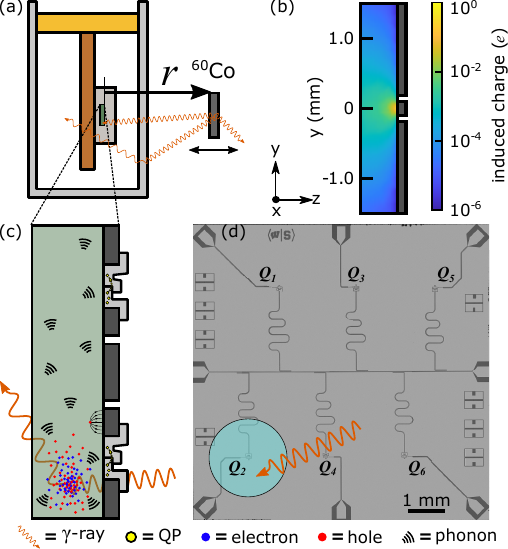}
 \caption{{\bf Experimental Setup.} (a) Schematic of dilution refrigerator, sample housing, and $^{60}$Co source. (b) Simulation of the induced offset charge on the qubit island, located at $y=0$ (not drawn to scale), for a unit charge at different locations in the substrate. (c) Chip schematic showing $\gamma$-ray impact, the ensuing charge cloud affecting the induced offset charge on the qubit island, and phonons leading to QP tunneling in the qubit electrodes. (d) Optical micrograph of chip with sketch of $\gamma$-ray impact and estimated charge-sensing area (light blue circle) for the qubit from simulations.}
\label{fig:intro}
\end{figure}

In our experiment we study two chips, each with nominally identical arrays of six charge-sensitive transmon qubits with a Nb ground plane [Fig.~\ref{fig:intro}(d)]. Both chips have an 8~mm $\times$ 8~mm Si substrate of 525-$\muup$m thickness. 
One chip has no back-side metallization (non-Cu) while the other has a 1-$\muup$m-thick Cu layer on the back side patterned into an array of (200-$\muup$m)$^2$ islands with a 50-$\muup$m spacing (1-$\muup$m Cu) for phonon downconversion. Details can be found in Appendix~\hyperref[app:experimental_details]{A} for device fabrication, parameters, and measurement setup.

We use a 100-$\muup$Ci $^{60}$Co source for active $\gamma$ irradiation. The dominant decay mode of $^{60}$Co is $\beta$ emission with a half-life of 5.27 years, which allows us to perform measurements over several months without significant changes in the source's activity. In this decay mode, $^{60}$Co undergoes $\beta$ decay into $^{60}$Ni and almost always emits two $\gamma$-rays with energies of 1.1732 MeV (99.85\% of the time) and 1.3325 MeV (99.98\% of the time). Other $\gamma$-rays are emitted with much lower probabilities and are negligible for these experiments~\cite{Browne2013}. The $\beta$ can also be neglected in our experiments as it cannot penetrate through the cryostat shielding. We position the source at the same height as the qubit chip, then adjust the $\gamma$-ray dose rate by varying the position $r$ of a small box housing the source along a line perpendicular to the surface of the chip [Fig.~\ref{fig:intro}(a), Fig.~\ref{fig:setup}]. Because each $\gamma$-ray is emitted in a random direction in a 4$\pi$ solid angle, we expect the dose rate to increase with $1/r^2$ along with the offset-charge jump rate and the QP poisoning rate from the impacts [see Fig.~\ref{fig:intro}(c) for schematic of processes].

Since the source is not collimated, we have a broad beam of $\gamma$-rays that allows for the contribution of secondary particles and $\gamma$-rays not initially directed at the sample [Fig.~\ref{fig:intro}(a)]. This allows us to not only have a higher effective dose rate for our sample from secondaries, but we can also position the source closer to the sample to further increase qubit poisoning because there is no collimator between the source and the vacuum jacket of the dilution refrigerator.
We use Geant4, a Monte Carlo-based high energy physics simulation toolkit, to model the trajectory of each $\gamma$-ray emitted from the source, including generation of secondary radiation from scattering in the various shielding layers of the cryostat and sample mount~\cite{Agostinelli2003,Allison2006,Allison2016}. The subsequent $e^{-}/h^{+}$ pair production, transport, scattering, and recombination are modeled with Geant4 Condensed Matter Physics (G4CMP)~\cite{Kelsey2023}. To capture how these free charges in the device substrate modify $\delta n_g$ of our qubits, we solve a variational form of Poisson's equation and calculate the fraction of field lines from a unit charge inside the substrate that terminate at the floating equipotential surface of the qubit island geometry using FEniCS [Fig.~\ref{fig:intro}(b)], an open-source finite element solver toolkit~\cite{Alnaes2015,Logg2012}. We also use a NaI scintillation detector to characterize the source and calibrate our modeling results (details in Appendix~\hyperref[app:source_char]{B} and \hyperref[app:simulations]{C}).

\section{Offset-Charge Jumps}
\label{sec:offset-charge}
The nonzero charge dispersion $\delta f$ for our qubits allows us to operate the qubits as sensors of their local charge environment such that we can detect nearby ionizing impacts. To measure the induced offset charge for a qubit, we use a standard charge-tomography sequence consisting of an $X/2$ pulse, an idle of $1/2\delta f$, followed by another $X/2$ pulse to give a qubit 1-state probability of $P_1 = \frac{1}{2}[d + \nu \cos(\pi \cos(2\pi n_g))]$, where $n_g = n_g^{\mathrm{ext}} + \delta n_g$, $n_g^{\mathrm{ext}}$ is the externally applied gate charge, and $\delta n_g$ is the environmental offset charge on the qubit island; $d$ and $\nu$ are fit parameters related to the qubit readout signal~\cite{Christensen2019, Wilen2021}. When we sweep the applied gate charge, we observe a characteristic periodic modulation of $P_1$; shifts of the modulation pattern along the gate charge axis can be used to extract $\delta n_g$. Thus, by quickly repeating scans, we can monitor $\delta n_g$ in time. We define a change in $\delta n_g$ that exceeds $0.15e$ to be an offset-charge jump (see Appendix \hyperref[app:offset_charge_jumps]{D.1.a} for details).

\begin{figure}[t!]
\centering
\includegraphics[width=8.6cm]{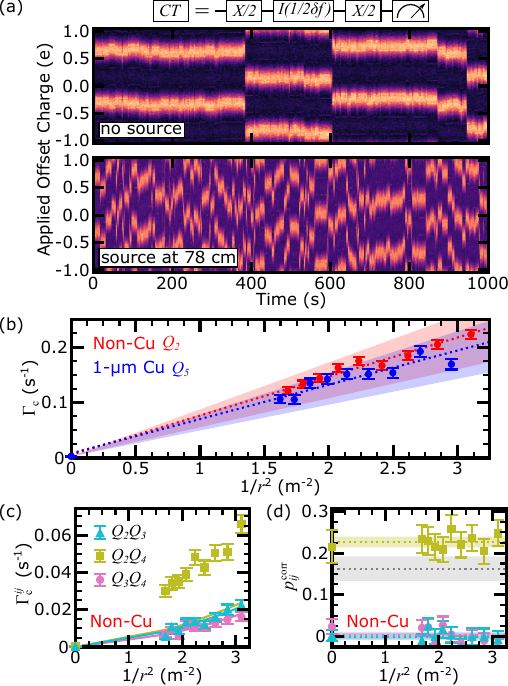}
 \caption{{\bf Measuring the charge environment.} (a) Charge tomography measurement sequence and example charge tomography scans repeated for 1000~s with no source in place (top) and with the source in place at $r=78$ cm (bottom). (b) Offset-charge jump rate for both devices across various source distances including the no-source rate at $1/r^2=0$. The dashed line is a linear fit and the shaded region is the simulation prediction for the non-Cu (1-$\muup$m Cu) data in red (blue). (c) Correlated jump rates between pairs of qubits on the non-Cu device for various source distances. The solid lines are the expected random background rates from the linear fit for the non-Cu single jump rates in (b) and the measurement window. (d) Correlation probabilities for the same qubit pairs shown in (c). The colored dotted line and shaded section represent the mean and standard deviation of the corresponding pair. The gray dotted line and shaded section represent the expected mean and standard deviation from the simulations for the $Q_2Q_4$ pair.}
\label{fig:charge}
\end{figure}

Figure~\ref{fig:charge}(a) shows examples of repeated charge tomography without and with the source present, showing a substantial increase in the rate of offset-charge jumps with the source. By collecting data for some time ($\sim$1 hour), we can quantify a charge jump rate. Figure \ref{fig:charge}(b) shows the offset-charge jump rate $\Gamma_{\rm c}$ for one example qubit on each chip for various source distances. We observe a clear $1/r^2$ dependence for both chips, as expected for uniform $\gamma$ emission from the source into a $4\pi$ solid angle. We also measure similar $\Gamma_{\rm c}$ for the same $r$ on both chips, confirming the presence of Cu back-side metallization does not affect the charge environment near the qubits. The slightly lower $\Gamma_{\rm c}$ on the 1-$\muup$m Cu chip is likely because it is physically situated on the opposite side of the cold finger from the source (Fig.~\ref{fig:setup}). Hence, there is more material shielding the chip from $\gamma$-rays, leading to slightly fewer impacts and a reduced rate of offset-charge jumps.

We also measure correlated offset-charge jumps between qubit pairs to help characterize the extent of the spread of electrons and holes from an impact site and to inform our numerical modeling. We use a staggered qubit control sequence (Appendix~\hyperref[app:measurement_setup]{A.3}) to perform charge tomography for up to four qubits at a time, allowing us to quantify correlated offset-charge jumps between qubit pairs and their dependence on distance between pairs. Figure \ref{fig:charge}(c) shows the correlated jump rates $\Gamma_{\rm c}^{ij}$ for three qubit pairs ($Q_iQ_j$) on the non-Cu chip across several source distances. Both the observed and random background rates are plotted, highlighting the elevated correlations for the closest pair of qubits ($Q_2 Q_4$, separated by 2.04~mm); at the same time, the two further pairs, $Q_2Q_3$ and $Q_3Q_4$, separated by 4.62 and 5.36~mm, respectively, do not exhibit excess correlated offset-charge jump rates above the expected random background. We also calculate the correlation probability $p_{ij}^{\rm corr}$, which represents the probability that an offset-charge jump on qubit $i$ coincides with a jump on qubit $j$~\cite{Wilen2021} (Appendix~\hyperref[app:offset_charge_jumps]{D.1.a}). We plot $p_{ij}^{\rm corr}$ for the same three qubit pairs across several source distances [Fig.~\ref{fig:charge}(d)]. Again, we observe an elevated correlation for the closest qubit pair, $p_{24}^{\rm corr} = 0.23(1)$, while the further separated pairs have correlation probabilities consistent with zero. We note that the length scale for correlated offset-charge jumps is somewhat larger than was observed in Ref.~\cite{Wilen2021}, but this is likely due to differences in the Si wafers used for the substrate and the qubit island geometry.

Following the approach of Ref.~\cite{Wilen2021}, we also model the charge dynamics from the controlled irradiation of our devices. We model the irradiation of the chip and subsequent ionizing impacts using Geant4, the relevant $e^{-}/h^{+}$ processes using G4CMP, and the induced charge on the qubit island using FEniCS. We can combine these simulations to generate simulated ionizing impacts and subsequent offset-charge shifts on all six qubit islands for each $^{60}$Co decay event. However, there are device-specific parameters that govern the mean free path of the charges: the charge trapping lengths $\Lambda_{\rm{trap}}^{e^-}$ and $\Lambda_{\rm{trap}}^{h^+}$ for electrons and holes, respectively, and the charge production efficiency $f_q$, following the definitions in Ref.~\cite{Wilen2021}. To determine these parameters, we use experimental results to inform our model: $p_{ij}^{\rm corr}$, the charge-shift asymmetry, and the ratio of the probabilities to observe an offset-charge jump and QP poisoning given a $\gamma$ impact occurred (further details in Appendix~\hyperref[app:Geant4/G4CMP]{C.3}).

With these simulation parameters determined, we can statistically characterize the charge-sensing area of our qubits to a typical charge burst from a $^{60}$Co deposition. We find that our qubit island geometry will sense an induced charge $\geq0.15e$ for any impact that is within $\sim$1~mm of the qubit island, shown as the light blue circle in Fig.~\ref{fig:intro}(d) (see Appendix~\hyperref[app:charge_foot]{C.4} for details). We also can accurately predict the bounds of the $\Gamma_{\rm c}$ slopes as shaded regions in Fig.~\ref{fig:charge}(b) for the non-Cu (red) and 1-$\muup$m Cu (blue) chips.
The estimated $p_{ij}^{\rm corr}$ values from our simulations for a close qubit pair ($Q_2Q_4$) are shown as a gray dashed line in Fig.~\ref{fig:charge}(d), and the 1-$\sigma$ uncertainty ranges are shown as the gray shaded region. The same prediction for the farther qubit pairs is consistent with zero, matching the experimentally observed values. More details on the suite of simulations used to make this prediction are included in Appendix \hyperref[app:simulations]{C}. The data used to fit the parameters is from a separate series of experiments where we conduct 4-way charge tomography measurements on the non-Cu device to have another close qubit pair ($Q_3Q_5$, with a separation of 2.04~mm) that we can analyze. We took data at several different source distances and for longer times ($\sim$30 hours) to improve counting statistics. The equivalent plot of $p_{ij}^{\rm corr}$ for this data is shown in Appendix~\hyperref[app:offset_charge_jumps]{D.1.a}.

\section{QP Parity Switching}

\begin{figure}[t!]
\centering
\includegraphics[width=8.6cm]{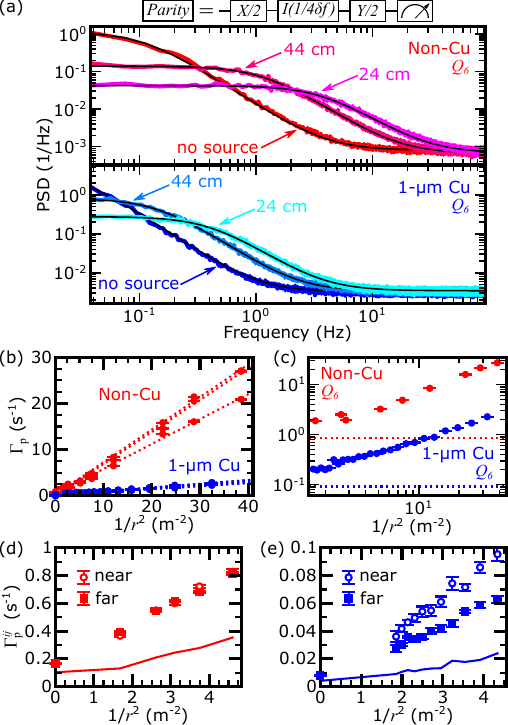}
 \caption{{\bf Measuring QP poisoning.} (a) Parity measurement sequence and example PSDs for $Q_6$ on the non-Cu (top) and 1-$\muup$m Cu (bottom) chips for different source-sample distances. (b) Characteristic parity switching rates for various source distances for the non-Cu (red) and 1-$\muup$m Cu (blue) devices including the no-source rate at $1/r^2=0$. The dashed lines are linear fits to the data for each qubit. (c) The same data from (b) plotted on a log scale and only for $Q_6$ on each device. The dashed horizontal lines are the no-source rates for $Q_6$ on each device. Average two-fold parity switching rates for (d) non-Cu, and (e) 1-$\muup$m Cu chips for qubit pairs near (open circles), and far (closed squares) from each other. The solid lines are the expected random background coincidence rate calculated from the single-qubit switching rates. Note the factor of 10 difference between the vertical axes.}
\label{fig:parity}
\end{figure}

\begin{figure*}[t!]
\centering \includegraphics[width=17.2cm]{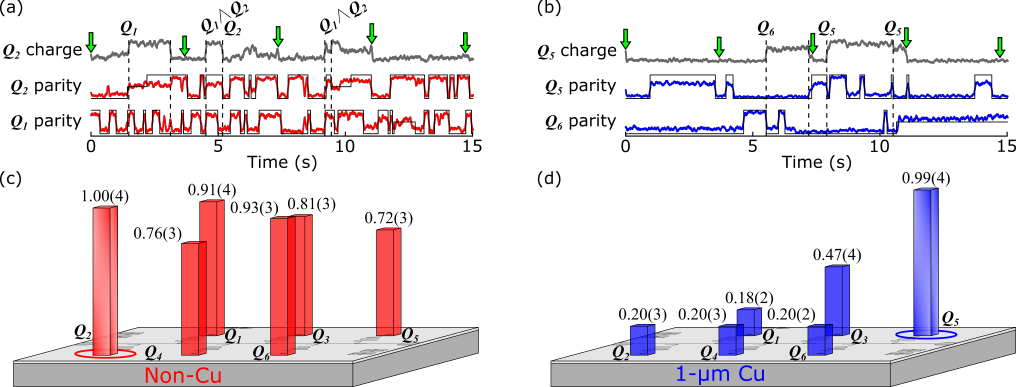}
 \caption{{\bf Correlating charge jumps and QP poisoning.} Example data traces showing the charge-sensing qubit's charge data in gray and QP parity data along with QP parity data for the qubit directly across the feedline for the (a) non-Cu chip in red, and the (b) 1-$\muup$m Cu chip in blue. The green arrows indicate the resetting of the voltage bias of the charge-sensing qubit. The dashed lines indicate charge jump detections. The labels above some charge jump detections indicate qubits that experienced a parity switch coincident with that detection. Poisoning probability distribution across the (c) non-Cu chip, and (d) 1-$\muup$m Cu chip, given that an offset-charge jump occurred on the charge-sensing qubit (circle at base, gradient shading). For all of the data shown, the source was in place at $r \approx 59$~cm.
\label{fig:probs}}
\end{figure*}

While an offset-charge jump indicates an ionizing impact near a qubit, we use measurements of the QP charge-parity switching rate $\Gamma_{\rm{p}}$ to characterize QP poisoning on each of the qubits, as an enhanced $\Gamma_{\rm{p}}$ indicates an elevated QP density in the qubit junction electrodes~\cite{Catelani2011, Marchegiani2022}. With the nonzero charge dispersion of our qubits, we use a standard parity-measurement sequence to map QP charge-parity onto $P_1$ for the qubit: we apply an $X/2$ pulse, idle for $1/4\delta f$, then apply a $Y/2$ pulse, followed by measurement of $P_1$~\cite{Riste2013, Serniak2019, Christensen2019}. We record single-shot parity measurements with a particular repetition time, between 250 $\muup$s and 10 ms depending on the $\gamma$ dose rate, to generate a time trace, to which we then apply a threshold from a 0/1 qubit readout calibration to produce a digital time trace. We compute the power spectral density (PSD) for this digital trace, which in general can be fit to a Lorentzian plus a white noise floor to extract the characteristic parity switching rate $\Gamma_{\rm{p}}$~\cite{Riste2013}. Figure \ref{fig:parity}(a) shows example parity-switching PSDs for one qubit on each chip with no source, as well as with the source present for two different distances. With no source present, we observe that $\Gamma_{\rm{p}}$, corresponding to the roll-off frequency of the PSD, is lower for the 1-$\muup$m Cu chip compared to the non-Cu chip due to the downconversion of pair-breaking phonons from background radioactive impacts by the back-side Cu islands; it remains lower when the source is introduced, even as both chips exhibit an increase in $\Gamma_{\rm{p}}$ with the source present. We quantify this increase through measurements of $\Gamma_{\rm{p}}$ vs. $1/r^2$ [Fig.~\ref{fig:parity}(b,c)]. Similar to the offset-charge jump measurements, we also observe a $1/r^2$ dependence, but now there is a dramatic difference between the two chips. For the same $r$, $\Gamma_{\rm{p}}$ for the 1-$\muup$m Cu chip is reduced by an order of magnitude compared to the non-Cu chip. We note that the faster measurement sequence for characterizing $\Gamma_{\rm{p}}$ compared to that for measuring $\Gamma_{\rm{c}}$ allows us to collect QP-parity switching data out to much closer source-sample distances (larger $1/r^2$).

We also track correlated parity switches between qubits for each chip. Figure~\ref{fig:parity}(d,e) show the average two-fold QP parity switching rate $\Gamma_{\rm p}^{ij}$ for neighboring qubit pairs $ij$ less than 3~mm apart (near) and all other pairs (far) for both devices with respect to $1/r^2$. Similar to $\Gamma_{\rm p}$, we observe $\Gamma_{\rm p}^{ij}$ for the 1-$\muup$m Cu chip is reduced by an order of magnitude compared to the non-Cu chip. Furthermore, we observe no separation dependence for the non-Cu chip,  consistent with the modeling in Ref.~\cite{Yelton2024}. Since there is no phonon downconversion layer for the non-Cu device, the phonons produced from $\gamma$-ray impacts can propagate throughout the substrate, impacting the qubit structures and causing QP tunneling with high probability. However, the 1-$\muup$m Cu device does exhibit a dependence on qubit separation, with the two-fold correlated parity switching for the near qubit pairs $\sim$50\% higher than for the far qubit pairs. 
%for the closest source distance. 
In this case, the Cu structures downconvert the phonon energy following a $\gamma$-ray impact and reduce the effective mean free path of the phonons.

\section{QP Poisoning Footprint}
\label{sec:QP_footprint}
In the previous two sections, we demonstrated that both $\Gamma_{\rm c}$ and $\Gamma_{\rm p}$ scale with $1/r^2$. Thus, by placing the $^{60}$Co source at a small $r$, we can ensure that the offset-charge dynamics and QP poisoning are dominated by $\gamma$-rays of known energy from the $^{60}$Co source rather than background radiation or cosmic-ray muons. We use this regime to study correlations between offset-charge jumps and QP poisoning. We designate one qubit in the array to be the \textit{charge-sensing qubit}. Based on our charge modeling described earlier, an offset-charge jump registered by this qubit gives us a charge trigger for a $\gamma$-ray impact within a radius of $\sim$1~mm of the qubit. We can then investigate the level of correlations in the QP poisoning for each of the qubits in the array following a particular impact.

For the charge-sensing qubit, we first apply the same charge tomography sequence described earlier to find the voltage bias corresponding to maximal charge dispersion. We then set the bias to this point and perform single-shot charge tomography measurements interleaved with single-shot QP parity measurements for all six qubits, including the charge-sensing qubit. Every 5000 single shots ($\sim$3.7 s), the voltage bias of the charge-sensing qubit is reset to account for any changes in the offset charge. While the method described in Sec.~\ref{sec:offset-charge} assigns changes in $\delta n_g$ greater than 0.15$e$ as offset-charge jumps, this protocol assigns steps in the single-shot charge data as offset-charge jumps (see Appendix~\hyperref[app:QP_poisoning_footprint]{D.1.c} for details). Although this approach makes it difficult to extract the precise magnitude of the offset-charge jump, it improves the timing precision for when the jump occurred, which is crucial for our coincidence analysis with the QP parity data.
 
When the charge-sensing qubit detects an offset-charge jump, we check the corresponding time window of the QP parity stream for each qubit and track coincident parity switches. From this approach, we obtain $p_i^{\rm obs}$, which is the ratio of observed coincidences to all occurrences of an offset-charge jump on the charge-sensing qubit during a fully measurable parity window for qubit $i$ (see Appendix~\hyperref[app:QP_parity_switching]{D.1.b} for details). For a given poisoning event, the QP tunneling rate is much faster than we can measure. Thus, we can only observe a change in parity for an odd number of switches, and the probability to observe a parity switch for a given event is 50\%.
To relate $p_i^{\rm obs}$ to a more useful poisoning probability $p_i^{\rm poison}$, we use the following equation
\begin{equation}
    p_i^{\rm poison} = 1 - \frac{1 - 2p_i^{\rm obs}}{1 - 2p_i^{\rm bkgd}}\,, 
\end{equation}
where $p_i^{\rm bkgd}$ is the expected probability for a random background coincidence for qubit $i$ with the offset-charge jump detected by the charge-sensing qubit; $p_i^{\rm bkgd}$ is determined by the single-qubit parity switching rate for qubit $i$, which is computed by a hidden Markov model (HMM) analysis, and the window size (see Appendix~\hyperref[app:QP_poisoning_footprint]{D.1.c} for details).

Figure~\ref{fig:probs}(a,b) shows example data traces for both devices for this protocol with the source in place at $r\approx59$~cm ($1/r^2\sim 2.9\,{\rm m}^{-2}$), where the offset-charge jumps are dominated by $\gamma$-ray impacts from $^{60}$Co [Fig.~\ref{fig:charge}(b)]. Steps in the charge data traces indicate offset-charge jumps, and the dashed lines indicate detection. The downward green arrows indicate when the applied bias is reset to the point of maximal charge dispersion for the charge-sensing qubit. The reset process takes $\sim$175~ms, but this time is omitted in the example traces for clarity. The solid black line over the parity data is the extracted digital signal from the HMM analysis. The qubit labels above the dashed detection lines indicate coincidences between the detected offset-charge jumps and parity switches for the labeled qubits. Figure~\ref{fig:probs}(c,d) shows $p_i^{\rm poison}$ for each qubit across each chip given that the charge-sensing qubit (circle at base, gradient shading) experienced an offset-charge jump. The example time traces shown in Fig.~\ref{fig:probs}(a,b) are brief segments from 12 (5)-hour experiments on the non-Cu (1-$\muup$m Cu) chip that are used to compute the $p_i^{\rm poison}$ levels in Fig.~\ref{fig:probs}(c,d).

The non-Cu chip shows a somewhat uniform distribution, with the charge-sensing qubit ($Q_2$), expected to be within $\sim$1~mm of the $\gamma$-ray impact from our modeling described earlier, heavily poisoned ($p_2^{\rm poison} = 1.00$) and the other qubits with slightly lower probabilities of being poisoned (0.72 - 0.93). This poisoning distribution is in good agreement with the numerical modeling of phonon and QP dynamics in Ref.~\cite{Yelton2024}, which exhibited a relatively uniform level of quasiparticle density $x_{\rm qp}$ following a $\gamma$-ray impact for a chip without phonon downconverting structures. The 1-$\muup$m Cu chip shows a quite different response. The charge-sensing qubit ($Q_5$) is still heavily poisoned, but the level of poisoning decreases significantly the farther each qubit is from the inferred impact site. For the nearest neighbor of the charge-sensing qubit, the correlated poisoning probability is reduced by $\sim$1/2, and by $\sim$1/5 for the other qubits that are further away. 

\begin{figure}[t!]
\centering
\includegraphics[width=8.6cm]{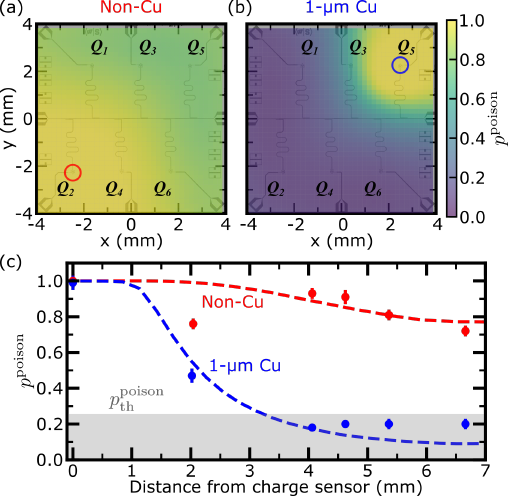}
 \caption{{\bf Modeling QP poisoning distribution.}
 Simulated poisoning probability distribution for the (a) non-Cu,  and (b) 1-$\muup$m Cu  chips. (c)  Calculated poisoning probability linecut at $y=x$ from modeling for both chips compared with experimental values plotted as a function of distance to the charge-sensing qubit. The shaded gray region represents poisoning levels below the threshold for robust error correction, as described in the text.}
\label{fig:modeled_footprint}
\end{figure}

We can estimate a QP poisoning footprint relevant to quantum error correction through the level of the resulting multi-qubit correlated poisoning. Robust error correction will require single-qubit errors below $10^{-4}$. Thus, the probability of two-qubit correlated errors should remain below $10^{-8}$~\cite{Wilen2021}. If we assume a surface code duty cycle of 1~MHz, the threshold two-qubit correlated error rate $r_{\rm th} = 10^{-2}$~s$^{-1}$. We can relate $r_{\rm th}$ to a threshold poisoning probability $p^{\mathrm{poison}}_{\mathrm{th}} = r_{\mathrm{th}}/R_\gamma$, where $R_\gamma$ is the $\gamma$-ray impact rate at the chip from background radiation. We estimate $R_\gamma$ by taking the average offset-charge jump rate with no source in place ($\sim$0.002~s$^{-1}$), measured by our charge-sensing qubits, and scaling this rate by the ratio of our charge-sensing area to the area of the entire chip, resulting in $R_\gamma \approx 0.04$~s$^{-1}$. Thus, our $p^{\mathrm{poison}}_{\mathrm{th}} \approx 0.25$. This value is much lower than the poisoning distribution on the non-Cu chip [Fig.~\ref{fig:probs}(c)], indicating that the QP poisoning footprint exceeds the chip dimensions. However, for the 1-$\muup$m Cu chip, only one pair of qubits (the charge-sensing qubit $Q_5$ and its nearest neighbor $Q_3$) exceeds this threshold [Fig.~\ref{fig:probs}(d)], indicating that the footprint is on the order of a few mm. We note that this $p_{\rm th}^{\rm poison}$ is specific to a particular $R_\gamma$ and is sensitive to its variation.

We compare these findings to a predicted QP poisoning footprint from applying the modeling results from Ref.~\cite{Yelton2024}. Using the predicted $x_{\rm qp}$ levels following a $\gamma$-ray impact from Ref.~\cite{Yelton2024}, we convolve with the radius of the charge-sensing area and assume a Poisson-distributed process to relate to $p^{\rm poison}$ (see details in Appendix~\hyperref[app:modeled_footprint]{C.5}). The resulting $p^{\rm poison}$ distribution is shown in Fig.~\ref{fig:modeled_footprint}(a,b). Figure~\ref{fig:modeled_footprint}(c) shows a diagonal linecut of the $p^{\rm poison}$ distribution compared against the $p_i^{\rm poison}$ values in Fig.~\ref{fig:probs}(c,d). Overall, we observe good agreement with the experimental values for both chips, with the exception of $Q_4$ on the non-Cu chip, which may have nearby defects leading to elevated phonon attenuation.

We expect a thicker Cu film, as in Ref.~\cite{Iaia2022}, should result in an even smaller poisoning footprint than our 1~$\muup$m-thick film. We also expect higher energy impacts, such as from cosmic ray muons, to produce larger poisoning footprints, consistent with measurements reported in Ref.~\cite{Harrington2024}. Note that this analysis is for $^{60}$Co $\gamma$-rays, which deposit on average $\sim$200~keV into the substrate (Appendix~\hyperref[app:Geant4]{C.1}), roughly twice the amount than from background radiation in a typical lab environment reported in Ref.~\cite{Wilen2021}. Further investigations are needed to establish a detailed picture of the energy dependence of the poisoning footprint. Ultimately, this framework can be applied for continued development of mitigation strategies, such as thicker back-side normal metal structures~\cite{Iaia2022} or low-gap superconducting films~\cite{Henriques2019, Karatsu2019, Yelton2024}, for further reductions in the poisoning distribution beyond the results demonstrated here.

%%%%%%%%%  Conclusion  %%%%%%%%% 
\section{Conclusion}

While most prior experimental characterizations of QP poisoning and correlated errors have relied on background radiation for random energy depositions, here we have demonstrated the controlled irradiation of $\gamma$-rays with known energy from a $^{60}$Co source. By monitoring offset-charge jumps on some qubits in the array, plus QP-parity switching on all of the qubits, we are able to characterize the QP poisoning footprint from $\gamma$-ray impacts. We observe a dramatic difference in the QP poisoning response to $\gamma$ irradiation and the poisoning footprint between chips with and without structures for mitigating pair-breaking phonons. We also connect our measured response to numerical simulations tracking $\gamma$-rays from our source and subsequent generation of secondaries for our system, as well as Monte Carlo modeling of the charge and phonon dynamics in our substrate geometry.

Characterizing QP poisoning footprints is crucial for the development of quantum error correction (QEC) schemes based on the identification of localized regions of transient error enhancements in the qubit lattice. Such QEC codes, combined with hardware mitigation of QP poisoning, can potentially help to reduce logical error rate and increase available computation time. As quantum processors grow to larger physical sizes with longer algorithmic runtime, incorporating radiation resilience into device designs will be vital. Further studies of active irradiation of qubit arrays with high-energy particles can be used to characterize the effectiveness of device designs and poisoning-mitigation strategies against correlated errors for quantum processors operating in radiation-exposed environments.

%%%%%%%%%  Acknowledgments  %%%%%%%%% 
\section{Acknowledgements}
This work is supported by the U.S. Government under ARO grant W911NF-22-1-0257. Fabrication was performed in part at the Cornell NanoScale Facility, a member of the National Nanotechnology Coordinate Infrastructure (NNCI), which is supported by the National Science Foundation (Grant NNCI-2025233). E.Y., C.P.L., and B.L.T.P acknowledge
partial support by the Laboratory for Physical Sciences
through Strategic Partnership Project EAOC0167012. K.D. acknowledges support by an appointment to the
Intelligence Community Postdoctoral Research Fellowship Program at Syracuse University, administered by
Oak Ridge Institute for Science and Education through
an interagency agreement between the U.S. Department
of Energy and the Office of the Director of National
Intelligence. The authors thank IARPA and Massachusetts Institute of Technology Lincoln Laboratory for providing the traveling-wave parametric amplifier (TWPA)~\cite{Macklin2015} used in this experiment. 
The authors also thank M. Soderberg for lending the NaI scintillator.

\begin{table*}[t!]
\begin{tabular}{ |p{1.5cm}||p{1cm}|p{1.5cm}|p{1.5cm}|p{1.5cm}|p{1.5cm}|p{1.5cm}|}
 \hline
 \multicolumn{7}{|c|}{Device Parameters} \\ \hline
 
 \hfil Device & \hfil Qubit & \hfil $f_{01}$ (GHz) & \hfil $f_{\rm{R}}$ (GHz) & \hfil $T_{1}$ ($\muup$s) & \hfil $\delta f$ (MHz) & \hfil $E_{\rm{J}}/E_{\rm{C}}$\\
 \hline

 \hfil\multirow{6}*{non-Cu}& \hfil $Q_1$ & \hfil 4.60 & \hfil 6.26 & \hfil 54(7) & \hfil 3.08 & \hfil 23 \\\cline{2-7}
                          &  \hfil $Q_2$ & \hfil 4.58 & \hfil 6.20 & \hfil 32(4) & \hfil 1.80 & \hfil 25 \\\cline{2-7}
                          &  \hfil $Q_3$ & \hfil 4.11 & \hfil 6.12 & \hfil 51(5) & \hfil 4.95 & \hfil 21 \\\cline{2-7}
                          &  \hfil $Q_4$ & \hfil 4.32 & \hfil 6.06 & \hfil 29(2) & \hfil 5.80 & \hfil 20 \\\cline{2-7}
                          &  \hfil $Q_5$ & \hfil 4.10 & \hfil 6.01 & \hfil 10(1) & \hfil 8.00 & \hfil 19 \\\cline{2-7}
                          &  \hfil $Q_6$ & \hfil 4.18 & \hfil 5.96 & \hfil 58(7) & \hfil 4.44 & \hfil 21 \\\cline{2-7}
 \hline

 \hfil\multirow{6}*{1-$\muup$m Cu}&  \hfil $Q_1$ & \hfil 4.62 & \hfil 6.26 & \hfil 18(3) & \hfil 2.78 & \hfil 24 \\\cline{2-7}
                          &  \hfil $Q_2$ & \hfil 4.30 & \hfil 6.20 & \hfil 24(2) & \hfil 14.8 & \hfil 17 \\\cline{2-7}
                          &  \hfil $Q_3$ & \hfil 4.30 & \hfil 6.13 & \hfil 24(4) & \hfil 3.85 & \hfil 22 \\\cline{2-7}
                          &  \hfil $Q_4$ & \hfil 4.36 & \hfil 6.06 & \hfil 17(3) & \hfil 5.77 & \hfil 21 \\\cline{2-7}
                          &  \hfil $Q_5$ & \hfil 4.34 & \hfil 6.03 & \hfil 19(2) & \hfil 5.40 & \hfil 21 \\\cline{2-7}
                          &  \hfil $Q_6$ & \hfil 4.32 & \hfil 5.98 & \hfil 22(3) & \hfil 3.51 & \hfil 22 \\\cline{2-7}
\hline
\end{tabular}
\caption{\textbf{Qubit and resonator parameters for the non-Cu and 1-$\muup$m Cu devices.}}
\label{tab:device_params}
\end{table*}

%%%%%%%%%  Appendix  %%%%%%%%%

\setcounter{section}{0}
\renewcommand{\thesubsection}{\arabic{subsection}}

\section*{Appendix A: Experimental and Device Details}
\label{app:experimental_details}
\setcounter{subsection}{0}

\subsection{Device fabrication}
\label{app:device_fab}

Both the non-Cu and 1-$\muup$m Cu chips are fabricated on high-resistivity ($>$10 k$\mathrm{\Omega}$-cm), 100-mm diameter,
525-$\muup$m thick Si wafers. The non-Cu wafer is single-side polished while the 1-$\muup$m Cu wafer is double-side polished for ease of the Cu island fabrication. Following the process described in Ref.~\cite{Iaia2022}, we use electroplating to deposit a 1-$\muup$m-thick Cu film on the back side of the wafer. We then pattern this film into $(200~\muup {\rm m})^2$ islands using partial cuts with a dicing saw. The ground planes of both devices are sputtered Nb films of 70-nm thickness. The ground planes are patterned as in Ref.~\cite{Iaia2022}, resulting in X-mon-style qubit islands with a 5-$\muup$m gap between the island and the ground plane. The Josephson junctions are made through shadow evaporation of Al to create electrode thicknesses of 40 and 80 nm, as was done in Ref.~\cite{Iaia2022}, and are designed to have $E_J/E_C \approx 20-25$. The wafers are then diced into individual chips for mounting in the sample packages.

\subsection{Device parameters}
\label{app:device_params}

The relevant parameters for both chips are listed in Table~\ref{tab:device_params}.

\subsection{Measurement setup}
\label{app:measurement_setup}

\begin{figure}[b]
\centering
 \includegraphics[width=8.6cm]{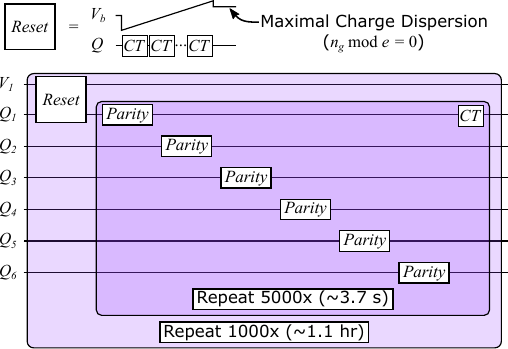}
 \caption{{\bf Staggered sequence of QP poisoning footprint measurement.} Pulse diagram of the QP footprint measurement showcasing the staggered sequence approach using $Q_1$ as the charge-sensing qubit. Note the pulse sequences for the $CT$ and $Parity$ blocks are defined in Fig.~\ref{fig:charge}(a) and Fig.~\ref{fig:parity}(a), respectively.}
\label{fig:staggered}
\end{figure}

For multi-qubit experiments, we implement a staggered measurement sequence (Fig.~\ref{fig:staggered}) %instead of the traditional multiplexing to reduce to 
to avoid crosstalk between closely spaced qubit frequencies (see Table~\ref{tab:device_params}) since the qubit rotations are driven through the common feedline. Thus, for our typical single-shot sequences, we measure each qubit sequentially with a wait time of 50-100 $\muup$s between each measurement and repeat for some number of single shots. This allows us to perform multi-qubit measurements with minimal crosstalk while only marginally increasing total measurement time when compared to a multiplexed sequence with simultaneous pulses. Moreover, since our moving average and coincidence window is $\sim$100 data points, this difference in the measurement times for each single shot is insignificant when performing our coincidence analysis. 

\begin{figure*}[thp]
\centering
 \includegraphics[width=17.2cm]{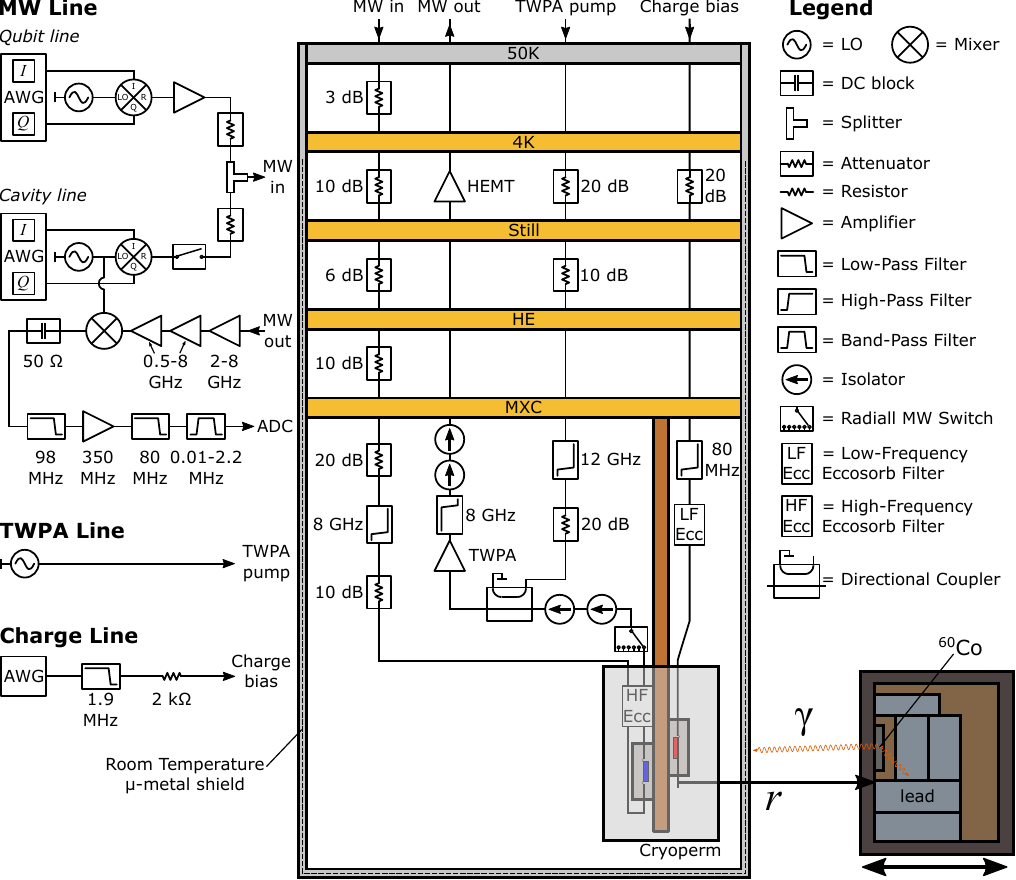}
 \caption{{\bf Experimental setup.} Wiring diagram for both room-temperature and cryogenic components. The MXC and Cryoperm shields have an IR-absorbent coating. The MW in and charge-bias lines are identical for the non-Cu (red) and 1-$\muup$m Cu (blue) devices, but only one for each is shown for 
 clarity. 
 %ease of viewing. 
 The bottom-right shows the $^{60}$Co disc source partially surrounded by lead bricks within a wooden box, which sits on top of a rollable cart (not shown). The height of the wooden box can be adjusted to align the source with the device being measured.}
\label{fig:setup}
\end{figure*}

Measurements for both chips are performed on the same cold finger and on the same cooldown in the dilution refrigerator. We use a standard heterodyne setup to conduct measurements, as shown in Fig.~\ref{fig:setup}, which details the attenuation, filtering, and shielding within the cryostat, as well as the room-temperature electronics. Also shown in Fig.~\ref{fig:setup}, the $\gamma$-ray source is located outside the cryostat, with lead bricks within a box on the opposite side of the source from the cryostat. This box sits on a rollable cart, and the height of the box is adjusted to align the source with the device being measured.

\section*{Appendix B: Background Radioactivity and Source Characterization}
\label{app:source_char}
\setcounter{subsection}{0}

Using a 1" NaI scintillation detector, we characterize the background radioactivity in our lab and from the $^{60}$Co source. Using the 2 photoabsorption peaks from the $^{60}$Co spectrum in Fig.~\ref{fig:Geant4}(e), we can calibrate our detector and verify that the source's activity exceeds the background radioactivity by a factor of $\sim$40 for a source-scintillator distance of 46 cm.

We also put the scintillator inside a nominally identical set of cryostat shielding to replicate the shielding environment for the chip in the experiment. By recording spectra at several distances with respect to the source and comparing it to the case with no shielding, we observe a 22\% higher event rate with the shielding from the excess contribution from secondaries.

\section*{Appendix C: Simulations}
\label{app:simulations}
\setcounter{subsection}{0}

\subsection{Geant4: NaI experiment model}
\label{app:Geant4}

\begin{figure}[h!]
\centering
 \includegraphics[width=\linewidth]{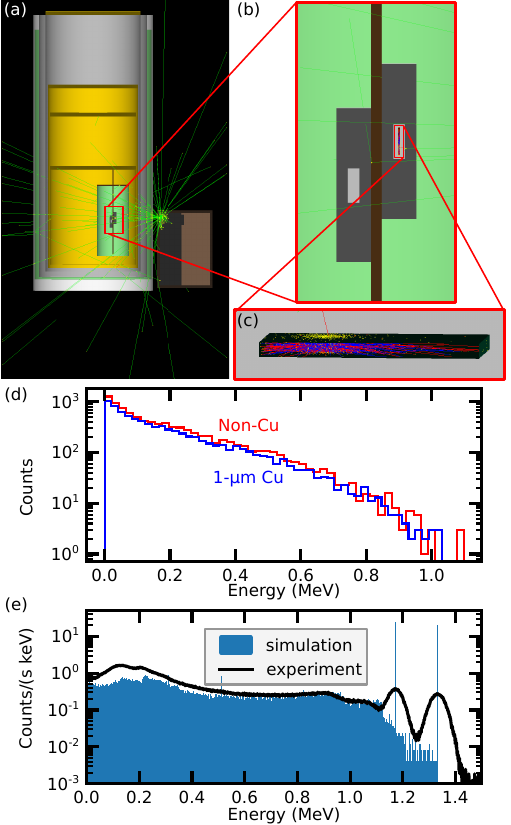}
 \caption{{\bf Geant4/G4CMP Simulation.} (a) Simulation of 100 $^{60}$Co decay events emitted in 4$\pi$ solid angle in the combined Geant4 and G4CMP simulation of the radiation source and cryostat vacuum jackets and heat shields. The particle trajectories of the emitted $\gamma$-rays are plotted in green, the trajectories of secondary $e^{-}$ particles are plotted in red. (b) Detail of the cryostat payload including both sample boxes. (c) We are only modeling the non-Cu Si crystal using G4CMP in this example. The produced electrons (blue) and holes (red) traverse the substrate geometry.  
 (d) Histogram of energy depositions from a simulation of 2 billion $^{60}$Co decay events for the non-Cu (red) and the 1-$\muup$m Cu (blue) samples with the source at the same distance from the face closest to the source. The $\gamma$-rays deposit an average energy of 192 (189) keV for the non-Cu (1-$\muup$m Cu) samples. (e) NaI scintillator spectrometer data (black), where the scintillator is encased in the cryostat heat shields and vacuum jackets and is 41~cm away from the $^{60}$Co source, is compared to a Geant4 simulation of this experiment (blue histogram).}
\label{fig:Geant4}
\end{figure}

We utilize two Monte Carlo-based High Energy Physics simulation software toolkits, namely Geant4 and Geant4 Condensed Matter Physics (G4CMP), to validate the ratio of offset-charge jump rates $\Gamma_{\rm c}$ [Fig.~\ref{fig:charge}(b)] to the QP charge-parity switching rates $\Gamma_{\rm p}$ [Fig.~\ref{fig:parity}(b)] observed in the experiments described in the main manuscript under $\gamma$ irradiation from a $^{60}$Co source~\cite{Agostinelli2003,Allison2006,Allison2016,Kelsey2023}.
To validate the model of the experimental setup, we start by simulating the NaI scintillator experiment described in the previous section using Geant4.
We model the lead-lined box containing the $^{60}$Co source, the cryostat vacuum jacket and heat shields, the sample package, and both the Si substrates of the devices [Fig.~\ref{fig:Geant4}(a)].
Note that for the modeling of the NaI calibration experiments, we did not include the Cu cold finger or the device and sample package, but included a 1" NaI material cylinder, matching the experimental arrangement.  

Each trial of the simulation models a $^{60}$Co decay, where a $\gamma$-ray of energy 1.3325~MeV is emitted and there is a 99.86\% probability of a $\gamma$-ray of energy 1.1732~MeV also being emitted; both particles are emitted in a random direction in a 4$\pi$ solid angle.   
The energy depositions within the NaI volume are recorded for multiple trials of simulated $^{60}$Co decays~[Fig.~\ref{fig:Geant4}(e)].
The Geant4 simulation does not capture the peak broadening in the measured spectrum due to the Gaussian statistics in converting scintillation photons to photoelectrons in the photomultiplier tube of the NaI detector \cite{Knoll10}.
Therefore, we validate the Geant4 modeling by focusing on the rate of photoabsorption events in the simulation and in the experimental spectrum to predict the known activity of the $^{60}$Co source. 
Specifically, we use the rate of the 1.3325~MeV $\gamma$-ray photoabsorptions because for every $^{60}$Co decay, a $\gamma$-ray of this energy is emitted; since it is the higher energy of the two, there is not a Compton edge contribution from a higher energy $\gamma$-ray, which is the case for the 1.1732~MeV $\gamma$-ray peak.

In a Geant4 simulation with 1 billion $^{60}$Co decay trials, with the $\gamma$ source at an identical distance as done in the experiment, we count 4905 (5865) photoabsorptions for the case with (without) the cryostat vacuum jackets and heat shields~[Fig.~\ref{fig:Geant4}(e)]. 
We fit a Gaussian to the corresponding photoabsorption peak in the experimental spectrum and assume that all counts on this peak are true photoabsorption events in the NaI detector.
We then sum the total number of counts on this peak and divide by the total measurement time of the experiment to find the rate of 1.3325~MeV $\gamma$ photoabsorptions; we measure 17.05(7)~s$^{-1}$ [21.26(8)~s$^{-1}$] for the experiment with [without] the vacuum jackets and heat shields. Dividing the measured rate by the probability of a photoabsorption from the Geant4 simulation, we arrive at an estimate of the source activity of 94(1)~$\muup$Ci [98(1)~$\muup$Ci], while the known source activity during the experiment was 96~$\muup$Ci, based on the manufacturer's calibration and the half-life of the source, indicating that our simplified geometric model of the cryostat captures the relevant rates of the $\gamma$-ray dynamics. 

We then take the geometry of the lead-lined box containing the source and the heat shields and vacuum jackets in the Geant4 modeling of the NaI experiments and use this in modeling the irradiation of our devices. To that end, we have included a Cu cold-finger, Al sample packaging, and Si device substrate in the simulation geometry. We model the substrate with G4CMP, an additional Monte Carlo toolkit designed to model $e^{-}/h^{+}$ pairs and phonons in cryogenic crystals \cite{Kelsey2023}. The $e^{-}/h^{+}$ pair production and transport within the substrate are modeled by default with the toolkit. However, there are experiment-specific parameters, such as the $e^{-}$ and $h^{+}$ trapping lengths ($\Lambda^{e^{-}}_{\textrm{trap}}$ and $\Lambda^{h^{+}}_{\textrm{trap}}$), which set the mean free path of the charges in the substrate and the efficiency for charges to not immediately recombine and transport throughout the substrate $f_q$, which we need to calibrate to experimental data. The latter parameter is included, as done in Ref.~\cite{Wilen2021}, to account for the absence of a static electric field throughout the device substrate. 

\subsection{FEniCS: Charge-sensitivity model}
\label{app:FEniCS}

\begin{figure}[h]
\centering
 \includegraphics[width=\linewidth]{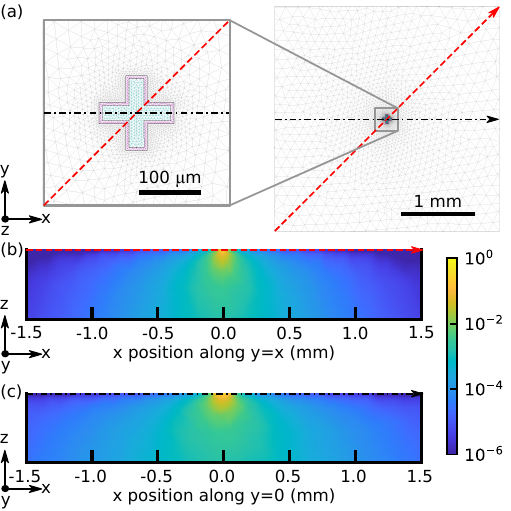}
 \caption{{\bf FEniCS induced charge simulation.} (a) The 2D surface mesh of the floating equipotential qubit island (cyan) and island gap (magenta) in the ground plane of the 3D mesh geometry for the 8~mm $\times$ 8~mm $\times$ 525~$\muup$m simulated Si substrate used in the FEniCS electrostatics simulations. Left shows the detail of the qubit island/gap geometry and the right shows the range of $x$ and $y$ positions where test charges were added into the substrate. The red dashed line indicates the linecut of the substrate cross-section shown in (b). Similarly, the dot-dashed black line indicates the substrate cross-section in (c). For each pixel in (b,c) a test charge is modeled and the color indicates the magnitude of the induced charge on the island. The colorbar corresponds to the unitless induced charge on the qubit island when normalized by the test charge. %\textcolor{red}{when normalized by $e$ (?)}.
\label{fig:Induced_Charge}}
\end{figure}

To connect the simulations to our experiments and determine realistic values of these three parameters, we must accurately model how free charges within the substrate modify the offset charge $\delta n_g$ of our charge-sensitive qubits. 
We calculate $\delta n_g$ by counting the fraction of field lines that terminate on the qubit island. 
The qubit island of our devices does not have the same azimuthal symmetry as the qubit islands of the devices used in Ref.~\cite{Wilen2021}.
Therefore, we model the electrostatic fields from a test charge within our device substrate in three dimensions by utilizing an open-source finite-element solver toolkit, FEniCS \cite{Alnaes2015,Logg2012}.
We solve a variational form of Poisson's equation for a 8-mm $\times$ 8-mm $\times$ 525-$\muup$m Si substrate, with a continuous ground plane and the qubit island and gap geometry at the center of the grounded surface~[Fig.~\ref{fig:Induced_Charge}(a)].
In a suite of simulations, we change the position of a unit test charge inside the substrate and calculate the fraction of field lines that terminate on the floating equipotential of the qubit island.
The results of this free-charge parameter sweep are shown in Fig.~\ref{fig:Induced_Charge}(b,c).

\subsection{Geant4/G4CMP: Charge parameter sweep}
\label{app:Geant4/G4CMP}

We then use the free-charge parameter sweep described in the previous section as an interpolated look-up table to compute 
%the induced offset charge 
$\delta n_g$ for each of the different qubits from the final positions of charges generated in the Geant4/G4CMP simulation. The boundaries of the substrate are modeled to be perfect absorbers of charge. Additionally, the charges are downsampled by a factor of ten for computational efficiency. This downsampling means that instead of, for example, 20,000 $e^{-}/h^{+}$ pairs being produced and tracked following a particular $\gamma$ impact, only 2,000 $e^{-}/h^{+}$ pairs are modeled for the same energy impact. To account for this, the magnitude of each charge is increased by a factor of ten when computing the induced charge on the qubit island. 
Note that when computing the simulated induced charge shifts we need to account for the 1$e$-periodicity of the charge tomography scan. Consequently, we can only measure offset-charge jump magnitudes $\leq\,$0.5$e$. 
Therefore, for a given impact in the simulation, the total induced charge from the simulation is aliased to an equivalent charge shift of magnitude $\leq\,$0.5$e$ to match the experimental data. Here we assume the charge production, transport, and trapping all occur between charge tomography measurements, implying that the final positions of charges contribute to the simulated shift in the offset-charge environment of the qubit. This assumption is valid given that the experimental repetition periods are $\sim$600~ms, much longer than the temporal dynamics of the charge bursts, where the average lifetime is on the order of tens of ns~\cite{Martinis2021}.

\begin{figure*}[ht!]
\centering
\includegraphics[width=\linewidth]{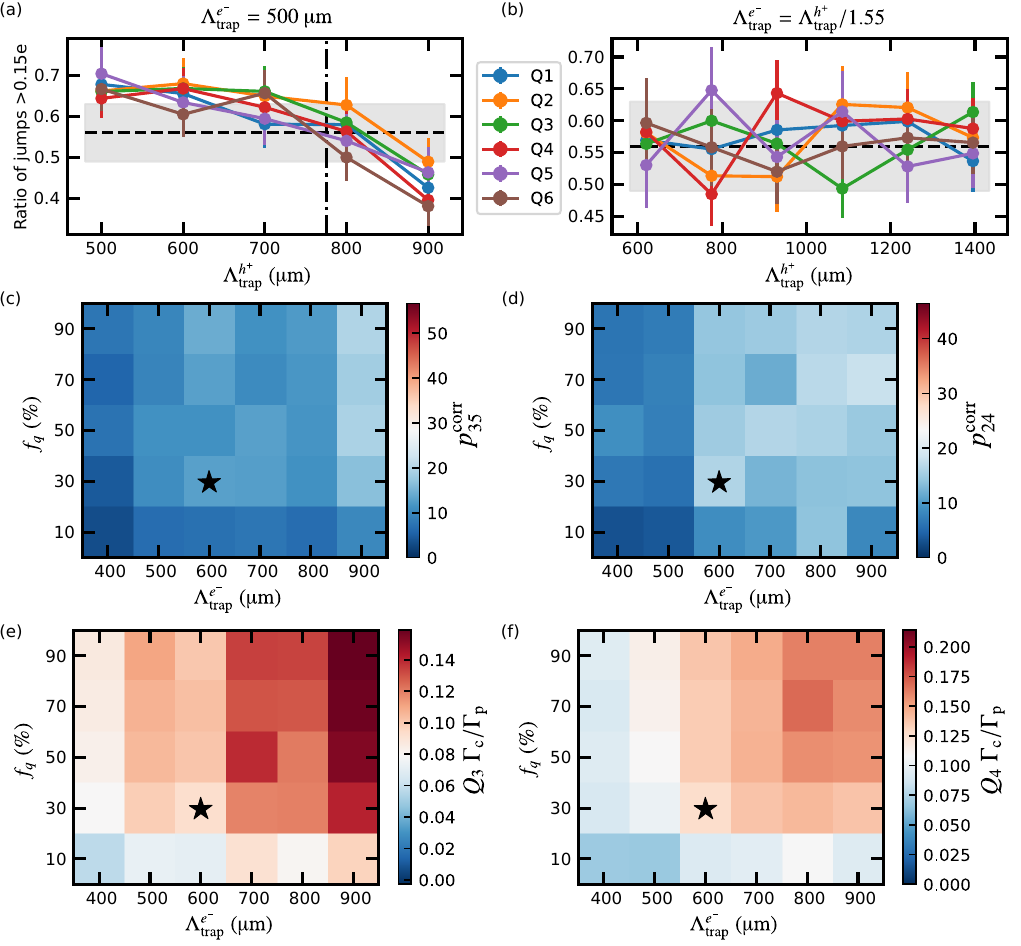}
\caption{{\bf Simulation charge parameters sweep.} Simulations of 20 million $^{60}$Co decays. Parts (a) and (b) show the asymmetry of charge jumps $>0.15e$ that are positive, the experimental average (dashed line) and standard deviation (gray area) are plotted in (a) and (b). (a) Sweep of the ratio of $\Lambda^{h^{+}}_{\textrm{trap}}$/$\Lambda^{e^{-}}_{\textrm{trap}}$ while keeping the $\Lambda^{e^{-}}_{\textrm{trap}}$ fixed at 500~$\muup$m. (b) Sweep of increasing electron and hole trapping lengths, [the lengths of which coincide with parts (c-f)], while keeping the ratio of $\Lambda^{h^{+}}_{\textrm{trap}}$/$\Lambda^{e^{-}}_{\textrm{trap}}$ constant (dot-dashed line in part (a)). The simulated correlation probability of (c) the $Q_3Q_5$ pair and (d) the pair $Q_2Q_4$ while sweeping the trapping lengths and $f_q$. The extracted $\Gamma_{\rm c}/\Gamma_{\rm p}$ ratio of (e) $Q_3$ and (f) $Q_4$. For parts (c-f), the white in the center of the color bar indicates the experimental target value. The black star indicates the simultaneous optimal trapping lengths and $f_q$ for all experimental quantities.}
\label{fig:charge_param_sweep}
\end{figure*}

Since we have a simulation protocol that mimics the output of the experimental data, we can sweep simulation parameters and calibrate to our experimental data. Generally, the electron and hole trapping lengths ($\Lambda^{e^{-}}_{\textrm{trap}}$ and $\Lambda^{h^{+}}_{\textrm{trap}}$) are not the same and depend on the density of acceptor and donor charge impurities within the substrate. We attribute an asymmetry in the sign of the measured charge shifts to a discrepancy between the transport of the two charge populations. 
Focusing on data from the non-Cu sample, we find that 56(7)$\%$ of charge shifts $>0.15e$ are positive. In our simulation protocol when we set $\Lambda^{h^{+}}_{\textrm{trap}}$/$\Lambda^{e^{-}}_{\textrm{trap}}=1$ [and $\Lambda^{e^{-}}_{\textrm{trap}}=500~\muup$m, see Fig.~\ref{fig:charge_param_sweep}(a)] we find that 67(5)$\%$ of charge shifts $>0.15e$ are positive. To match our experimental measurements, we increase the hole-trapping length in a suite of simulations since excess holes located closer to the island will induce negative offset-charge shifts [Fig.~\ref{fig:charge_param_sweep}(a)]. We find that a charge-trapping length ratio of $\Lambda^{h^{+}}_{\textrm{trap}}$/$\Lambda^{e^{-}}_{\textrm{trap}}=1.55$ [dot-dashed line in Fig.~\ref{fig:charge_param_sweep}(a)] agrees with our experimentally measured charge-shift asymmetry [gray region and dashed line in Fig.~\ref{fig:charge_param_sweep}(a,b)]. 
We next use this charge-trapping length ratio as a constraint in our parameter search. At the same time, we still need to know the magnitude of the charge-trapping lengths and the efficiency for charges to not immediately recombine, $f_{q}$, which we can determine using other experimental quantities.

The probability of a correlated offset-charge shift on a pair of qubits has been shown to depend on the geometrical spacing of the two qubits in Ref.~\cite{Wilen2021}. We observe a similar dependence, as shown in our measurements [Fig.~\ref{fig:charge}(d)]. This correlation probability depends on both the qubit island design and the transport of charges from an impact site, making this quantity useful to compare the simulation results to experiments. Thus, we focus on the correlation probability of two pairs of qubits separated by $\sim$2~mm ($Q_2Q_4$ and $Q_3Q_5$). We can produce a correlated charge-jump probability from our simulation data, following the analysis done in Ref.~\cite{Wilen2021}. In Fig.~\ref{fig:charge_param_sweep}(c,d) we sweep $f_q$ and the electron trapping length $\Lambda^{e^{-}}_{\textrm{trap}}$, while maintaining $\Lambda^{h^{+}}_{\textrm{trap}}$/$\Lambda^{e^{-}}_{\textrm{trap}}=1.55$. We simulate 20 million $^{60}$Co decay events for each simulation with the $\gamma$-rays emitted in a 4$\pi$ solid angle. The correlation probability is independent of radiation flux [Fig.~\ref{fig:charge}(d)], therefore, we simulate the source being much closer to the cryostat than what is done experimentally [Fig.~\ref{fig:Geant4}(a)]. This allows for more ionizing events per simulated $^{60}$Co decay, increasing the computational efficiency of the simulation runs. 

From the same suite of simulations we can also extract a simulation prediction for the ratio of $\Gamma_{\rm c}/\Gamma_{\rm p}$, that is, the ratio of the slopes in Fig.~\ref{fig:charge}(b) and Fig.~\ref{fig:parity}(b), leveraging the probabilities found for the non-Cu chip in Fig.~\ref{fig:probs}(c). From the statistics in Fig.~\ref{fig:probs}(c), we know that for a particle impact detected by the charge-sensing qubit, we see a nearly uniform response on all the qubits, with a 
poisoning probability near unity and $\sim50\%$ probability of a QP charge-parity switch being observed. %\textcolor{red}{(*we mean "being observed" rather than "occurring" right?)}. 
Taking the average QP charge-parity switching probability of all qubits, we assume the probability of a QP charge-parity switch occurring given a particle impact occurred is 45(1)$\%$. From the simulation of the charge transport and the charge sensing of the qubit island, we can simulate the probability of a charge jump $>0.15e$ occurring given a particle impact occurred. The ratio of these two probabilities is equivalent to the ratio $\Gamma_{\rm c}/\Gamma_{\rm p}$. From the simulations, we use simple counting statistics of offset-charge shifts $>0.15e$ to find the probability of an offset-charge shift given a particle impact occurred on the chip. The results are shown in Fig.~\ref{fig:charge_param_sweep}(e,f). We find that $\Lambda^{h^{+}}_{\textrm{trap}}=930~\muup$m, $\Lambda^{e^{-}}_{\textrm{trap}}=600~\muup$m, and $f_q=30\%$ have the best agreement for the multiple experimental quantities we compare in Fig.~\ref{fig:charge_param_sweep}(c-f) (denoted by a star in Fig.~\ref{fig:charge_param_sweep}(c-f)). These parameters are used in the simulation results reported in the main manuscript.

The estimate for the correlation probability of the $Q_3Q_4$ qubit pair (spaced by $\sim$4.6~mm) is near zero, consistent with experimental values. However, we did not use the statistics of this qubit pair in our parameter search optimization (Fig.~\ref{fig:charge_param_sweep}) given that we would need to run impractically long simulations to arrive at reasonable counting statistics to determine the event probabilities for this qubit pair with a large separation. Nevertheless, we still report the prediction using optimized simulation parameters, and the agreement with the experiments indicates that our simulations capture the physics of our experiments reasonably well.

Additionally, in the main manuscript, we report bounds on the slopes of the $\Gamma_{\rm c}$ versus $1/r^2$ slopes~[Fig.~\ref{fig:charge}(b)]. These bounds were calculated by taking the fitted slopes from the $\Gamma_{\rm p}$ versus $1/r^2$ data for the non-Cu data [Fig.~\ref{fig:parity}(b)] and for each of these slopes multiplying by the $\Gamma_{\rm c}/\Gamma_{\rm p}$ estimate, identical to the calculations done in Fig.~\ref{fig:charge_param_sweep}(c,d). Given the spatial dependence of the QP charge-parity switching probabilities from the location of the particle impact [Fig.~\ref{fig:probs}(d)] for the 1-$\muup$m Cu chip, we cannot treat the probability of a parity switch given a particle impact as being uniform across the chip, as done for the non-Cu case. Therefore the bounds are defined based on the results of the non-Cu sample and we account for the relative $\gamma$ flux due to the mounting position of this sample by using the hit rate from Geant4 simulation of 2 billion $^{60}$Co decays, with the source located at the same distance from the substrate for both samples [Fig.~\ref{fig:Geant4}(d)]. Thus, the difference in slopes in Fig.~\ref{fig:charge}(b) can be explained by the cryostat configuration relative to the $^{60}$Co source.

\subsection{Charge-sensing footprint}
\label{app:charge_foot}

Given the simulation parameters we have determined in the previous section, we can use these results to quantify the typical charge distribution after a particle impact. We create a characteristic charge burst from the resulting charge dynamics of 100 simulated $^{60}$Co $\gamma$-ray energy depositions on the device. For the charges of all 100 impacts, we set the initial $x$ and $y$ positions to $x=y=0$~mm, while the initial $z$ position remains unchanged from the original simulation. This allows us to statistically sample the possible initial $z$-positions from the energy depositions, while we have forced the charges to have been initialized at $x=y=0$~mm. The result is a large burst of charge that is initialized on a line through the substrate. This burst of charge is approximately 100 times bigger than a typical $\gamma$ energy deposition. To account for this we reduce the simulated charge of the charge carriers by a factor of 100.

\begin{figure}[h!]
\centering
\includegraphics[width=2.9in]{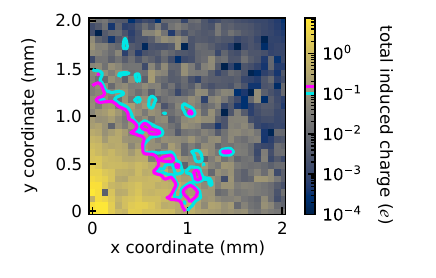}
\caption{{\bf Simulated charge-sensing footprint.} 
Characteristic charge burst from multiple $^{60}$Co $\gamma$-ray energy depositions initialized at $x=y=0$. At each pixel coordinate, a qubit island, the same as in Fig.~\ref{fig:Induced_Charge}(a), is located and the total induced charge on the island determines the color of each pixel. The magenta (cyan) line indicates the contour where the total induced charge is $0.15e$ ($0.1e$).}
\label{fig:charge_footprint}
\end{figure}

Now that we have this characteristic charge burst from a simulated $^{60}$Co $\gamma$-ray, we can sweep the location of the qubit island and calculate the induced charge on the island using the results presented in Fig.~\ref{fig:Induced_Charge}(b,c). Note that it is computationally more efficient to fix the large data store of the charge positions and sweep the location of the qubit island, rather than the reverse.
Each pixel in Fig.~\ref{fig:charge_footprint} represents a simulated $x$- and $y$-coordinate position of a qubit island relative to the characteristic charge burst at $x=y=0.$ The coloring of each pixel represents the total induced charge on the island. In this case, we are not aliasing the induced charge to be $\pm 0.5e$ as we did in matching results to experimental data. The magenta (cyan) line in Fig.~\ref{fig:charge_footprint}  represents the contour where the total induced charge is $0.15e$ ($0.1e$) on the qubit island. There is an inherent roughness to the contours that we attribute to the discretized formulation of the characteristic charge burst, that is, we use the weighted burst of multiple impacts rather than fitting a smooth function to the distribution of multiple charge bursts. Therefore, the shape of the contour in Fig~\ref{fig:charge_footprint} is not general but is likely characteristic of a true charge burst, which would not have been captured by treating the burst as a smooth distribution in space. The average distance of the $0.15e$ ($0.1e$) contour from the initial position of the charge burst at $x=y=0$ is 1060~$\muup$m (1200~$\muup$m). We use the distance from the $0.15e$ contour as the radius to define the characteristic charge-sensing area for detecting energy depositions from a $^{60}$Co source using the offset-charge jumps for our qubit islands. However, the distance from the $0.1e$ contour is still relevant as the method of determining offset-charge jumps for the QP poisoning footprint measurement is sensitive to jumps as low as $\sim$$0.1e$ [Fig.~\ref{fig:jumpDetection}(b)].

\subsection{QP poisoning footprint}
\label{app:modeled_footprint}

Given that we have determined the charge-sensing footprint, we can apply the modeling results presented in Ref.~\cite{Yelton2024} to the data presented in Fig.~\ref{fig:probs} to determine a predicted QP poisoning footprint. The G4CMP simulations in Ref.~\cite{Yelton2024} used controlled phonon-injection experiments to determine realistic phonon boundary conditions using devices that are nominally identical to the ones used in this work. Additionally, this work modeled the $x_{\rm qp}$ generation in the simulated qubit electrodes of a hypothetical device that consists of a dense array of qubits spaced by 200~$\muup$m  due to a $\gamma$-ray impact. A $\gamma$-ray energy deposition of 100~keV was modeled, which is a factor of $\sim$2 lower than the average energy deposited from the $\gamma$-rays from the $^{60}$Co source.

We can connect these simulations to the experiments by simulating energy depositions of 191.6 (188.9)~keV [Fig.~\ref{fig:Geant4}(d)] for the non-Cu (1-$\muup$m Cu) devices. The energy depositions are modeled to occur in the corner of the device, similar to both charge sensor locations used in the Fig.~\ref{fig:probs} data. We then take the $x_{\rm qp}$ values on the dense array at a time that coincides with the largest poisoning footprint, as defined in Ref.~\cite{Yelton2024}, at $\sim$$5~\muup$s after the impact. This largest poisoning footprint is then convolved on a $1$-mm radius circle to simulate the average $x_{\rm qp}$ response from any single-point charge jump triggered in the experimental data presented in Fig.~\ref{fig:probs}. We then relate this averaged $x_{\rm qp}$ to the poisoning probability $p^{\rm poison}.$ We assume the probability for $n$ QP tunneling events to occur given some normalized quasiparticle density $\lambda = x_{\rm qp}/\overline{x_{\rm qp}}$, where $\overline{x_{\rm qp}}$ is a characteristic $x_{\rm qp}$ threshold, in the qubit junction electrodes is Poisson distributed $p(n{\rm \text{-}tunnel~events})=\lambda^n e^{-\lambda}/n!.$ Therefore, the probability for any number of tunneling events to occur is $p^{\rm poison}=1-e^{-\lambda}.$ We then fit $\overline{x_{\rm qp}}$ using the experimental $p_i^{\rm poison}$ in Fig.~\ref{fig:probs}(c,d). Figure~\ref{fig:modeled_footprint}(c,d) shows the resulting $p^{\rm poison}$ distribution for both chips with the convolved impact centered on each respective charge-sensing qubit. We then take a diagonal linecut ($y=x$) and plot $p^{\rm poison}$ with respect to the distance from the charge-sensing qubit.

\section*{Appendix D: Measurement and Analysis Details}
\setcounter{subsection}{0}

\subsection{Correlated events}
\label{app:correlated_events}

 Our measurements allow for the tracking of correlated parity switching events, correlated offset-charge jump events, and correlations between both offset-charge and parity events. Each measurement has been implemented as a staggered sequence (see Appendix~\hyperref[app:measurement_setup]{A.3}), which involves a wait time (50-100 $\muup$s) between the readout of each qubit. This wait time is insignificant to our coincident tracking because the window size to determine coincident events ranges from 53 - 628~ms, depending on the source distance and the type of measurement (offset-charge, QP parity, or QP poisoning footprint).
 %\textcolor{red}{(*what do we mean by measurement here?)}
 For our offset-charge measurements without the source present, we increased the number of averages to improve our signal-to-noise ratio resulting in a window size of $\sim$10 s, further minimizing the effect of the staggered measurement times. Note that this longer window size is still short enough to measure our observed no-source $\Gamma_{\rm c}$, which range from 0.0011(1) s$^{-1}$ to 0.0030(3) s$^{-1}$.

\subsubsection*{\textbf{a. Offset-charge jumps}}
\label{app:offset_charge_jumps}
\begin{figure}[h!]
\centering
\includegraphics[width=8.6cm]{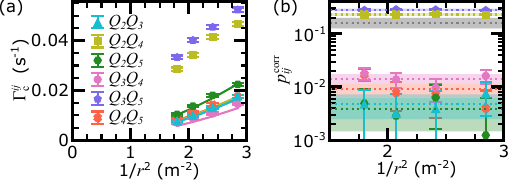}
 \caption{{\bf Measuring the charge environment.} (a) Correlated jump rates between pairs of qubits on the non-Cu device for various source distances. (b) Correlated probabilities for the same qubit pairs shown in (a). The colored dotted line and shaded section represent the mean and standard deviation of the corresponding pair. The gray dotted line and shaded section represent the expected mean and standard deviation from the simulations for the $Q_2Q_4$ pair.}
\label{fig:charge_4way}
\end{figure}

To understand correlated offset-charge jumps for different qubit pairs, we perform staggered charge tomography for 3 or 4 qubits on the non-Cu device. In order to track offset-charge jumps, we fit each repeated time slice to obtain the change in offset-charge between slices $\Delta q$. This measurement aliases large $\Delta q$ to be between $-0.5e$ and $0.5e$. We define an offset-charge jump to be $|\Delta q| > 0.15e$. We observe that a threshold of $0.15e$ is less sensitive to the exact choice of threshold compared to $0.1e$, which was used in Ref.~\cite{Wilen2021}. Next, we count the number of correlated jumps between different qubit pairs and calculate the correlation probability $p_{ij}^{\mathrm{corr}}$ between qubits $i$ and $j$ with the following equation described in Ref.~\cite{Wilen2021}:
\begin{equation}
    p_{ij}^{\rm corr} = \frac{2p_{ij}}{p_i^{\rm obs} + p_j^{\rm obs}}\,,
\end{equation}
where $p_{ij}$ is the true probability of a two-fold correlated offset-charge jump and is represented by the equation
\begin{equation}
    p_{ij} = \frac{p_{ij}^{\rm obs} - p_i^{\rm obs}p_j^{\rm obs}}{1+p_{ij}^{\rm obs} - (p_i^{\rm obs} + p_j^{\rm obs})}\ .
\end{equation}
For the data presented in Fig.~\ref{fig:charge}(d), we observe $p_{24}^{\rm corr} = 0.23(1)$ for the closest qubit pair, $Q_2 Q_4$, with a separation of $\sim$2 mm, while the other pairs with larger separations have $p_{ij}^{\rm corr}$ consistent with zero. In an effort to inform our model with another close qubit pair as a fitting parameter, we connected a fourth charge line to bias $Q_5$ on the non-Cu device, increasing our number of unique qubit pairs and adding an additional close pair, $Q_3 Q_5$. We also performed longer measurements ($\sim$30 hours) for improved statistics. Measurements were taken at several different source distances and can be seen in Fig.~\ref{fig:charge_4way}. These measurements were taken several months after the collection of the data presented in Fig.~\ref{fig:charge}(b,c), leading to slightly lower rates because of the small decrease in source activity. We observe that the lower rates have no effect on $p_{ij}^{\rm corr}$, where $p_{24}^{\rm corr}$ is 0.232(4), consistent with the value from the data shown in Fig.~\ref{fig:charge}(d). For the other close qubit pair, $p_{35}^{\rm corr} = 0.283(4)$. Both of these values are used as targets for the numerical simulation of the charge dynamics.

We also observe slightly nonzero charge correlation probabilities for the farther qubit pairs. From Fig.~\ref{fig:charge_4way}(b) we measure a weak distance dependence for the far qubit pairs, but are still limited by counting statistics due to the rarity of these events. The separation for the far pairs are as follows: $Q_3Q_4\textrm{ (pink)}=4.62$~mm, $Q_4Q_5\textrm{ (orange)}=5.36$~mm, $Q_2Q_3\textrm{ (light blue)}=5.36$~mm, and $Q_2Q_5\textrm{ (green)}=6.70$~mm. This suggests the presence of occasional charge shifts for a particular impact sensed by far qubit pairs, but these are far less prominent than events sensed by nearest neighbor qubits.

\subsubsection*{\textbf{b. QP parity switching}}
\label{app:QP_parity_switching}

Since the offset charge is not actively stabilized, the process for tracking coincident parity switches starts with determining when the charge has jumped to the degeneracy point, and the parity readout fidelity vanishes. To determine this, we first apply a moving average to the data. We then define a \textit{masking window} of the averaged data with length $1/(\Gamma_{\rm{p}} \Delta t)$, where $\Delta t$ is the repetition period. We then sweep through the entire data stream and fit a normal distribution for each masking window $i$. For a masking window to be accepted (unmasked), it must satisfy at least one of the following two conditions. First, the separation between the mean $\mu_i$ of the window $i$ and the mean of the entire data stream $\bar\mu$ is greater than the standard deviation $\sigma_i$ of the window,
\begin{equation}
    |\mu_i - \bar\mu| > \sigma_i\ .
\end{equation}
This accepts data with optimal parity readout fidelity, since the window of data is sufficiently above or below the average of the dataset $\bar\mu$, indicating that the parity data is mapped to the 0 or the 1 state. Note that this analysis assumes that there is an equal probability for the data to be in the odd or even parity state. This condition alone would mask sections of data that experience switches that cross over $\bar\mu$, which would ignore the switching events we want our data analysis protocol to track. So, the second condition handles this by also accepting data where
\begin{equation}
    \sigma_i > \alpha \bar\sigma\,,
\end{equation}
where $\bar\sigma$ is the standard deviation of the entire data stream. We choose $\alpha$ to be dependent on the readout of the qubit, more specifically, the state-mapping fidelity $F$ of the PSD fit to the parity-switching dataset \cite{Riste2013}. As defined in Ref.~\cite{Riste2013}, the probability for a state detection error is $(1-F)/2$. Due to imperfect readout fidelity and occasional charge jumps, we assume that these processes manifest as additive Gaussian white noise with a standard deviation of $\sigma$. Assuming this noise contribution is identical in the mapping to both the $\ket{0}$ and the $\ket{1}$ states, we can then relate the probability of a state detection error as the overlap of two Gaussians, with identical standard deviation $\sigma$ and means separated by $s$, the $\ket{0}$-$\ket{1}$ state separation along the qubit readout signal axis. This calculation gives the following relationship: $(1-F)/2=\left[1+{\rm erf}(s/2\!\sqrt{2}\sigma)\right]/2$. In general, the standard deviation of data that is distributed by two Gaussians with equal standard deviations and different means separated by $s$ is $\sigma_{\rm total}^2 = \sigma^2+\left(s/2\right)^2$. %$\bar\sigma = \sqrt{\sigma^2_{\rm noise}+\left(s/2\right)^2}$.
We then use this result to compute the relationship between $\sigma_i$ and $\bar\sigma$, which we equate to $\alpha$. Thus,
\begin{equation}
    \alpha = \sqrt{\frac{1 + \frac{1}{4}s^2}{1 + 2(\mathrm{erf}^{-1}(F)\sqrt{n})^2}}\,,
\end{equation}
where $s$ is our state-separation in units of $\sigma$ and $n$ is the number of data points in our moving average. Lastly, we choose to mask all data with a state detection error $>5\%$, which corresponds to $s = 3.29$. For our datasets with $n = 40$, this results in $\alpha$ between 0.24 and 0.82. This masking approach is compatible with our prior masking technique detailed in Ref.~\cite{Iaia2022}.

Once masked, we use a Hidden Markov model (HMM) to identify the parity states and convert the data to a digital signal. We sweep a \textit{coincidence window} (53 - 93 ms with source, 400 - 413 ms without source) equal to $n \Delta t$ to count coincident parity switches between qubits. Because $n$ is constant across all parity measurements (with the exception of the non-Cu data with no source present to match the coincidence window in Ref.~\cite{Iaia2022}), this window remains constant in data points, but changes its length in time depending on the repetition period we use. Measurements taken where the QP parity is switching faster (source is closer) require a faster repetition period and result in shorter window times. Further details of this process can be found in Ref.~\cite{Iaia2022}.

\subsubsection*{\textbf{c. QP poisoning footprint}}
\label{app:QP_poisoning_footprint}

\begin{figure}[hb!]
\centering
\includegraphics[width=8.6cm]{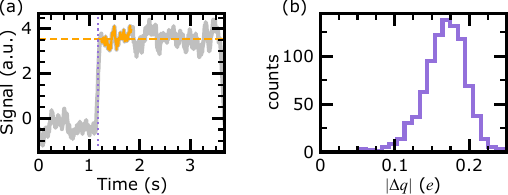}
 \caption{{\bf Detecting and estimating magnitude of offset-charge jumps for QP poisoning footprint measurement.} (a) Example jump and detection. The gray line is the averaged single-point charge data in between two charge resets. The orange line is the subset of the single-point charge data that is averaged to determine the signal level (orange dashed line) after the jump. The purple dotted line corresponds to the time of the detected jump identified by our routine. (b) Histogram of detected offset-charge jump magnitudes for a subset of the data obtained from the QP poisoning footprint measurement for the non-Cu device.}
\label{fig:jumpDetection}
\end{figure}

\begin{table*}[t!]
\begin{tabular}{|m{1.2cm}||m{2.1cm}|m{2.1cm}|m{2.1cm}|m{2.1cm}|m{2.1cm}|m{2.1cm}|}
 \hline
 \multicolumn{7}{|c|}{Correlated Charge and Parity Rates}\\
 \hline
 \hfil Device & \hfil Qubit & \hfil Counts & \hfil Unmasked & \hfil $p_{i}^{\rm obs}$ & \hfil $p_{i}^{\rm bkgd}$ & \hfil $p_{i}^{\rm poison}$ \\
 \hline
 \hfil\multirow{6}*{non-Cu} & \hfil $Q_1$ & \hfil 1351 & \hfil 2889 & \hfil 0.47(1) & \hfil 0.1539(6) & \hfil 0.91(4) \\\cline{2-7}
                & \hfil $\boldsymbol{Q_2}$ & \hfil \textbf{1306} & \hfil \textbf{2621} & \hfil \textbf{0.50(1)} & \hfil \textbf{0.1288(5)} & \hfil \textbf{1.00(4)} \\\cline{2-7}
                & \hfil $Q_3$ & \hfil 1342 & \hfil 3150 & \hfil 0.43(1) & \hfil 0.1173(5) & \hfil 0.81(3) \\\cline{2-7}
                & \hfil $Q_4$ & \hfil 1053 & \hfil 2593 & \hfil 0.41(1) & \hfil 0.1101(5) & \hfil 0.76(3) \\\cline{2-7}
                & \hfil $Q_5$ & \hfil 904  & \hfil 2344 & \hfil 0.39(1) & \hfil 0.0981(6) & \hfil 0.72(3) \\\cline{2-7}
                & \hfil $Q_6$ & \hfil 1433 & \hfil 3014 & \hfil 0.48(1) & \hfil 0.1355(6) & \hfil 0.93(3) \\\cline{2-7}
 \hline \hline
 \hfil\multirow{6}*{1-$\muup$m Cu} & \hfil $Q_1$ & \hfil 123 & \hfil 1149 & \hfil 0.11(1) & \hfil 0.0199(3) & \hfil 0.18(2) \\\cline{2-7}
                & \hfil $Q_2$ & \hfil 219 & \hfil 1398 & \hfil 0.16(1) & \hfil 0.0730(6) & \hfil 0.20(3) \\\cline{2-7}
                & \hfil $Q_3$ & \hfil 248 & \hfil 978  & \hfil 0.25(2) & \hfil 0.0361(5) & \hfil 0.47(4) \\\cline{2-7}
                & \hfil $Q_4$ & \hfil 160 & \hfil 1208 & \hfil 0.13(1) & \hfil 0.0380(4) & \hfil 0.20(3) \\\cline{2-7}
                & \hfil $\boldsymbol{Q_5}$ & \hfil \textbf{718} & \hfil \textbf{1445} & \hfil \textbf{0.50(2)} & \hfil \textbf{0.0241(3)} & \hfil \textbf{0.99(4)} \\\cline{2-7}
                & \hfil $Q_6$ & \hfil 145 & \hfil 1238 & \hfil 0.12(1) & \hfil 0.0238(3) & \hfil 0.20(2) \\\cline{2-7}

 \hline
\end{tabular}
\caption{\textbf{Probabilities of seeing a parity switch given a charge jump occurred.} The data for the charge-sensing qubit on each chip is in bold. The duration of the datasets were 11.78 and 5.35 hours for the non-Cu and 1-$\muup$m Cu devices, respectively.}
\label{tab:probs}
\end{table*}

The QP poisoning footprint measurement allows for the offset-charge to be monitored on the charge-sensing qubit while also measuring parity for all six qubits (Fig.~\ref{fig:probs}). The charge data for the charge-sensing qubit is a time trace with the same repetition period as the parity time traces. This allows for direct coincidence tracking between offset-charge jumps and parity switches. The process of detecting an offset-charge jump involves identifying a step in the charge time trace. To do this, we first apply a moving average of 100 data points to the charge data. We next convolve the averaged charge data with a step function of width 200 data points. We then assign peaks in the convolution past a certain threshold as offset-charge jumps. Figure~\ref{fig:jumpDetection}(a) shows an example detected jump. To estimate the jump size, we average the signal value after the jump, and relate this level to the most recent full charge tomography scan when the bias was reset. This allows us to translate our signal value to several possible offset-charge values (because the charge tomography response function is periodic in offset charge). We then assign the magnitude of the charge jump to be the shortest distance between the found offset-charge value and the bias point we set for the corresponding charge-bias reset. This means that we alias all magnitudes to be $<0.25e$. A histogram of detected offset-charge jump magnitudes for the non-Cu data shown in Fig.~\ref{fig:probs}(a,c) is plotted in Fig.~\ref{fig:jumpDetection}(b). 
The shape of the curve and the lack of jumps $<0.1e$ are consequences of the single-point charge scan used for these QP footprint measurements. In this case, the offset charge is biased near the point of maximal charge dispersion, where there is minimal sensitivity to small charge jumps.

Once we have a list of offset-charge jump occurrences, we analyze the parity streams with the same approach as described in Appendix~\hyperref[app:QP_parity_switching]{D.1.b} with a moving average and coincidence window size of 100 data points to match the charge data. Because we collect this data for much longer than for the parity measurements, we can afford to mask off more data to increase the quality of accepted data and still retain sufficient statistics. Thus, we choose to mask data with state detection error $>1\%$, which corresponds to $s = 4.65$.
This results in $\alpha$ between 0.25 and 0.43, with the exception of $Q_2$ on the 1-$\muup$m Cu device where we intentionally mask off more data with an $\alpha = 0.5$ because of its large $\delta f$ (see Table~\ref{tab:device_params}). We then examine the parity data stream for each qubit within the window size for all offset-charge jump indices. If the parity stream is fully unmasked within the coincidence window of the offset-charge jump, it is registered as \textit{unmasked}. If any or all of the stream is masked in the examined area, the offset-charge jump is excluded from our statistics. If the qubit is unmasked and experiences a parity switch, it is registered as a \textit{count}. The number of counts and unmasked segments for each qubit for the data shown in Fig.~\ref{fig:probs} can be seen in Table~\ref{tab:probs}.

\begin{figure}[hb!]
\centering
 \includegraphics[width=\linewidth]{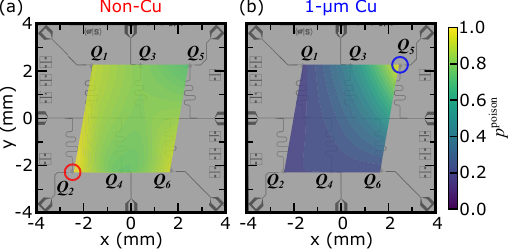}
 \caption{{\bf QP poisoning map.}
 Interpolated contour map based on the $p_i^{\rm poison}$ values shown in Fig.~\ref{fig:probs}(c,d) for the (a) non-Cu and (b) 1-$\muup$m Cu chip. The charge-sensing qubit for each chip is circled.}
\label{fig:QP_footprint}
\end{figure}

The observed probability $p_i^{\rm obs}$ is the ratio of the number of counts to unmasked segments for qubit $i$. The background probability $p_i^{\rm bkgd}$ is the expected probability for a random coincidence between a parity switch for qubit $i$ with the detected offset-charge jump on the charge-sensing qubit. This background probability is the product of the single-qubit parity switching rate determined by the HMM analysis and the coincidence window size. Next, we derive an expression using $p_i^{\rm bkgd}$ and $p_i^{\rm obs}$ to compute a more intuitive quantity for characterizing QP poisoning, the poisoning probability $p_i^{\rm poison}$, which is the quantity plotted in Fig.~\ref{fig:probs}(c,d). When we observe a parity switch on qubit $i$ coincident with an offset-charge jump detected by the charge-sensing qubit, there are two potential underlying processes: (1) a poisoning event of $Q_i$ occurred directly related to the $\gamma$-ray impact registered by the charge-sensing qubit with probability $p_i^{\rm poison}$, (2) an unrelated poisoning event of $Q_i$ occurred with probability $2p_i^{\rm bkgd}$ (the factor of 2 is included since we can only observe an odd number of parity switches in an event when determining $p_i^{\rm bkgd}$). We define $p_i^{\rm obs}$ by finding the complement of these events:
\begin{equation}
    p_i^{\rm obs} = \frac{1}{2}\left[1-\left(1-p_i^{\rm poison}\right)\left(1-2p_i^{\rm bkgd}\right)\right]\ .
\end{equation}
Solving for the poisoning probability, we find
\begin{equation}
    p_i^{\rm poison} = 1 - \frac{1-2p_i^{\rm obs}}{1-2p_i^{\rm bkgd}}\ .
\end{equation}
Note that $p_i^{\rm poison}$ is linear with respect to $p_i^{\rm obs}$. When $p_i^{\rm obs} = 0.5$, $p_i^{\rm poison} = 1$, and when $p_i^{\rm obs} = p_i^{\rm bkgd}$, $p_i^{\rm poison} = 0$.

We also perform a cubic interpolation of the six qubit locations and $p_i^{\rm poison}$ values to obtain a poisoning map in between the qubits (see Fig.~\ref{fig:QP_footprint}). This exhibits reasonable qualitative agreement with the simulated poisoning probability shown in Fig.~\ref{fig:modeled_footprint}.

\subsection{$T_1$ measurements}

\begin{figure}[b]
\centering
 \includegraphics[width=\linewidth]{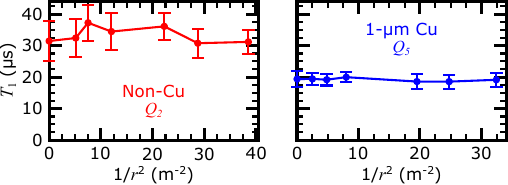}
 \caption{{\bf Measuring $T_1$ vs $1/r^2$.} Measured $T_1$ with respect to source distance for both non-Cu (red) and 1-$\muup$m Cu (blue) devices. 
\label{fig:T1}}
\end{figure}

In addition to the offset-charge jump and QP parity switching rates as a function of source distance as detailed in the main paper, we also characterized $T_1$ with a conventional measurement sequence consisting of a $\pi$-pulse followed by a variable delay before qubit readout (Fig.~\ref{fig:T1}). The lack of correlation between $T_1$ and source distance is explained by relatively few points in our $T_1$ measurement sequence being affected by $\gamma$ impacts. For a given $T_1$ measurement, we take $\sim$200 delay points and $\sim$200 averages at a repetition period between 150 - 500 $\muup$s, resulting in a minimum of 2000 measurement points per second. At the closest source distance, $r\approx17$~cm for the non-Cu chip, we observe $\sim$25 parity switches per second. Because we only see poisoning events with an odd number of parity switches, we expect there to be $\sim$50 QP poisoning events per second for our closest source distance. From the poisoning measurements performed in Refs.~\cite{Iaia2022} and \cite{Yelton2024}, we expect the recovery timescale following a $\gamma$ impact for our chip design and packaging to be $\sim$100 $\muup$s. This is short enough such that a particular $\gamma$ impact will only have an effect on a single point in the $T_1$ measurement sequence on average. Thus, with the poisoning rate (50~s$^{-1}$) only being a small fraction of our measurement rate (2000~s$^{-1}$), this measurement of $T_1$ should be insensitive to the source even for our closest distance, consistent with our observations. 

Although there is no measurable dependence on $1/r^2$, we observe slightly reduced $T_1$ times for the 1-$\muup$m Cu chip when compared to the non-Cu chip (see Table~\ref{tab:device_params}). 
While breaking up the back-side Cu film into islands eliminates the severe reduction in $T_1$ that would otherwise result from the transmission-line mode that would be formed, some metallic loss in the Cu islands still couples weakly to the electric field oscillations of the qubit, causing a slight reduction in $T_1$~\cite{Martinis2021}. This issue could be addressed while retaining the enhanced phonon downconversion by instead using a low-gap superconductor, as modeled in Ref.~\cite{Yelton2024}.

\bibliography{main}

\end{document}